\documentclass[aps,twocolumn,showpacs,preprintnumbers,nofootinbib,prd,superscriptaddress,10pt]{revtex4-2}

\usepackage{graphicx,amssymb,amsmath,amsthm,amsfonts,epsfig,epsf,fixmath}
\usepackage[usenames]{color}
\usepackage{epstopdf}
\usepackage{subfigure}
\usepackage{isotope}
\usepackage{physics}
\usepackage{bbold}
\usepackage{aas_macros}
\usepackage{bm}
\usepackage{comment}
\usepackage{dcolumn}
\usepackage{enumitem}
\usepackage{latexsym}
\usepackage{rotating}
\usepackage{longtable}

\usepackage{enumerate}
\usepackage{tensor,multirow}
\usepackage{url}
\usepackage[linktocpage]{hyperref}
\usepackage{soul}

\newcommand{\tn}{\textnormal}

\begin{document}

%%%%%%%%%%%%%%%%%%%%%%%%%%%%%%%%%%%%%%%%%%%%%%%%%%%%%%%
\title{Relativistic Corrections to the CBF Effective Nuclear Hamiltonian}
%%%%%%%%%%%%%%%%%%%%%%%%%%%%%%%%%%%%%%%%%%%%%%%%%%%%%%%
\author{Andrea Sabatucci}
\altaffiliation[Present address: ]{INFN, Sezione di Pisa, 56127 Pisa, Italy}
\affiliation{Dipartimento di Fisica, ``Sapienza'' University of Rome, 00185 Roma, Italy}
\address{INFN, Sezione di Roma, 00185 Roma, Italy}
%%%%%%%%%%%%%%%%%%%%%%%%%%%%%%%%%%%%%%%%%%%%%%%%%%%%%%%
\author{Omar Benhar}
\affiliation{INFN, Sezione di Roma, 00185 Roma, Italy}
%\affiliation{Dipartimento di Fisica, ``Sapienza'' University of Rome, Piazzale A. Moro, 5. 00185 Roma, Italy}
%%%%%%%%%%%%%%%%%%%%%%%%%%%%%%%%%%%%%%%%%%%%%%%%%%%%%%%
\author{Alessandro Lovato}
\affiliation{Physics Division, Argonne National Laboratory. Argonne, Illinois 60439, USA}
\address{INFN, Trento Institute of Fundamental Physics and Applications, 38123 Trento, Italy}
%%%%%%%%%%%%%%%%%%%%%%%%%%%%%%%%%%%%%%%%%%%%%%%%%%%%%%%

%%%%%%%%%%%%%%%%%%%%%%%%%%%%%%%%%%%%%%%%%%%%%%%%%%%%%%%
\begin{abstract}
We discuss the inclusion of relativistic boost corrections into the CBF effective nuclear Hamiltonian, derived from a realistic model of two- and three-nucleon interactions using the formalism of correlated 
basis functions and the cluster expansion technique. Different procedures to take into account the effects of boost interactions are compared on the basis of the ability to reproduce the nuclear matter equation 
of state obtained from accurate quantum many-body calculations. The results of our study show that the  
repulsive contribution of the boost interaction significantly depends on the underlying model 
of the non relativistic potential. On the other hand, the dominant relativistic correction 
turns out to be the corresponding reduction of the strength of repulsive three-nucleon interactions, leading to a significant softening of the equation of state of nuclear matter at supranuclear densities.

\end{abstract}
%%%%%%%%%%%%%%%%%%%%%%%%%%%%%%%%%%%%%%%%%%%%%%%%%%%%%%%
\maketitle

%%%%%%%%%%%%%%%%%%%%%%%%%%%%%%%%%%%%%%%%%%%%%%%%%%%%%%%
\section{Introduction}
\label{intro}
%%%%%%%%%%%%%%%%%%%%%%%%%%%%%%%%%%%%%%%%%%%%%%%%%%%%%%%

According to the paradigm of quantum many-body theory, all nuclear systems\textemdash from the deuteron
to neutron stars\textemdash can be treated as collections of
point-like protons and neutrons interacting via instantaneous potentials~\cite{bob}. It should be kept in mind, however, that this
description, while proving adequate to explain a wealth of nuclear properties, is inherently inconsistent with the requirement of causality, 
because it may lead to predict a speed of sound larger
then the speed of light in dense nuclear matter.

Pioneering studies of relativistic effects in both nuclear matter and the three-nucleon bound states were carried
out by Coester {\it et al.} in the 1970s and 1980s; see Refs.~\cite{Coester:NM,Coester:NNN}. In 1990 Brockmann and Machleidt reported the results of a thorough analysis of nuclear matter properties 
based on the one-boson-exchange model of nuclear dynamics~\cite{PhysRevC.42.1965}. The relativistic  
correction to the energy, which turned out to be repulsive, was obtained using the Hartree-Fock approximation and the Dirac-Bruekner formalism; see Ref.~\cite{Brockmann:1996xy}.

More recently,
relativistic corrections to the properties of the three- and four-nucleon systems have been thoroughly
analysed by the authors of Ref.~\cite{Forest_Arriaga} using a relativistic Hamiltonian\textemdash obtained from
state-of-the-art phenomenological potentials
following the procedure described in Ref.~\cite{boost}\textemdash and Quantum Monte Carlo techniques. The results of this study indicate that, owing to the occurrence of
large cancellations, the dominant relativistic effects, referred to as
boost corrections, are driven by the total momentum of the interacting particles. Boost corrections to the ground-state energies turn out to be
repulsive, and account for a sizeable fraction of the potential energy originating from irreducible repulsive interactions involving three nucleons. Very recently, similar conclusions have been also reached by the authors of Refs.~\cite{Yang:2022esu,Yang:2024wsg} using neural-network quantum states.

Relativistic boost corrections to the energies of pure neutron matter (PNM) and isospin-symmetric nuclear matter (SNM), computed within the
scheme described in Refs.~\cite{boost,Forest_Arriaga}, are included in the equation of state (EOS) of charge-neutral $\beta$-stable matter derived
by Akmal, Pandharipande and Ravenhall~\cite{APR}, widely employed in studies of neutron star properties.
The results of Ref.~\cite{APR} were obtained from accurate variational calculations performed with a highly-realistic phenomenological nuclear Hamiltonian comprising the Argonne $v_{18}$ (AV18) nucleon-nucleon (NN) potential~\cite{AV18} and the Urbana IX (UIX) three-nucleon (NNN) potential~\cite{UIX_2,UIX_1}. They show that the suppression of repulsive three-nucleon interactions associated with relativistic boost
corrections leads to a conspicuous softening of the EOS at supranuclear density, which appreciably affects the mass and radius
of stable neutron stars obtained from the solution of the equations of Tolman, Oppenheimer and Volkoff. The resulting maximum mass
turns out to decrease from 2.3 to 2.2~$M_\odot$, with $M_\odot$ being the solar mass, while the radius of a 1.4 $M_\odot$ neutron star is reduced from 12.7 to 11.6 Km.
The inclusion of boost corrections also alleviates the problem of violation of causality, pushing its occurrence towards higher density.

Over the past decade, the dynamical model underlying  the work of Ref.~\cite{APR} has been used,
in conjunction with the formalism of Correlated Basis Function (CBF)~\cite{FF:CBF} and the
cluster expansion technique~\cite{CLARK197989}, to derive a density-dependent effective interaction suitable to carry out perturbative
calculations of nuclear matter properties at both zero and non-zero 
temperature~\cite{BL:2017,Benhar_Lovato_Camelio,lucas_temperature,Benhar:Universe}.
The resulting NN potential includes the contributions of two- and three-nucleon forces, and is properly renormalised to take into account
the screening of non-perturbative interactions arising from strong short-range correlations among the nucleons.

The CBF effective Hamiltonian is designed to reproduce the EOSs of PNM and SNM obtained from accurate many-body calculations, carried out using a 
simplified version of the  AV18+UIX Hamiltonian  
without taking into account the relativistic boost corrections considered in Ref~\cite{APR}. 
In this article, we discuss an extension of the approach of Ref.~\cite{BL:2017} allowing to include these corrections, which are known to affect significantly the properties
of nuclear matter in the high-density regime relevant to neutron stars.

The rest of this article is organised as follows.
The phenomenological nuclear Hamiltonian providing the basis of the CBF effective interaction and the associated
relativistic boost corrections are discussed in Section~\ref{dynamics}, while Section~\ref{renormalisation} is devoted to a short description
of the procedure employed to derive the nuclear Hamiltonian of Ref.~\cite{BL:2017}, which essentially 
amounts to a renormalisation of nuclear forces in matter. The impact of relativistic boost corrections  
on the energy per particle of nuclear matter and the explicit expression of the underlying effective interaction are analysed in Sections~\ref{results} and \ref{boost:veff}, 
respectively. Finally,  in Section~\ref{outlook} we summarise our findings and state the conclusions.

%%%%%%%%%%%%%%%%%%%%%%%%%%%%%%%%%%%%%%%%%%%%%%%%%%%%%%%
\section{Dynamical model}
\label{dynamics}
%%%%%%%%%%%%%%%%%%%%%%%%%%%%%%%%%%%%%%%%%%%%%%%%%%%%%%%

In this section, we outline the model of nuclear dynamics underlying our approach, as well as the formalism employed to include the contribution of relativistic boost interactions.

%%%%%%%%%%%%%%%%%%%%%%%%%%%%%%%%%%%%%%%%%%%%%%%%%%%%%%%
\subsection{Nuclear Hamiltonian}
\label{nuclear:hamiltonian}
%%%%%%%%%%%%%%%%%%%%%%%%%%%%%%%%%%%%%%%%%%%%%%%%%%%%%%%

In non relativistic many-body theory, nuclear dynamics are described by the Hamiltonian
\begin{align}
H_{\rm NR} = \sum_{i=1}^{A} \frac{{\bf p}_i^2}{2m} + \sum_{j>i=1}^{A} v_{ij}
 + \sum_{k>j>i=1}^A V_{ijk} \ ,
\label{H:A}
\end{align}
where $A$ is the number of particles, ${\bf p}_i$ and $m$ denote the momentum of the $i$-th nucleon and its
mass, and the potentials $v_{ij}$ and $V_{ijk}$ describe NN and NNN interactions, respectively.

Phenomenological NN potentials are determined by fitting the observed properties of the
two-nucleon system\textemdash including the deuteron binding energy, magnetic moment and electric quadrupole
moment, as well as the data obtained from the measured NN scattering cross sections\textemdash and
reduce to Yukawa's one-pion-exchange potential at large distance. They are usually written in the form
\begin{align}
v_{ij}=\sum_p v^{p}(r_{ij}) O^{p}_{ij} \ ,
\label{eq:NN_1}
\end{align}
where the functions $v^p$ only depend upon the distance between the interacting particles, $r_{ij} = |{\bf r}_i - {\bf r}_j|$. The operators
$O^{p}_{ij}$ with $p =1,\ldots,6$ account for the strong spin-isospin dependence of nuclear forces, as well as for the occurrence of non-central interactions. They are defined as  
\begin{align}
O^{p \leq 6}_{ij} = [1, (\boldsymbol{\sigma}_{i}\cdot\boldsymbol{\sigma}_{j}), S_{ij}]
\otimes[1,(\boldsymbol{\tau}_{i}\cdot\boldsymbol{\tau}_{j})]  \ ,
\label{av18:2}
\end{align}
where $\boldsymbol{\sigma}_{i}$ and $\boldsymbol{\tau}_{i}$ are Pauli matrices acting in spin and isospin space, respectively, and 
\begin{align}
S_{ij}=\frac{3}{r_{ij}^2}
(\boldsymbol{\sigma}_{i}\cdot{\bf r}_{ij}) (\boldsymbol{\sigma}_{j}\cdot{\bf r}_{ij})
 - (\boldsymbol{\sigma}_{i}\cdot\boldsymbol{\sigma}_{j}) \
 \label{S12}
\end{align}
is the tensor operator. 
Highly-realistic models, such as the Argonne $v_{18}$ (AV18) potential~\cite{AV18}, include twelve additional terms. The contributions corresponding to $p~=~7,\ldots,14$ are associated with non-static components of the potential,  including spin-orbit and other angular momentum-dependent terms, while those corresponding to $p=15,\ldots,18$ take into account small violations of charge symmetry
and charge independence.

Three-body forces are long known to be required to model the interactions of extended composite bodies, such as protons and neutrons, without
considering their internal structure explicitly; see, e.g., Ref.~\cite{Friar:3BF}. In nuclear many-body theory, the inclusion of {\it irreducible} NNN forces,  described by the potential $V_{ijk}$,  is needed to explain both the observed binding energies of the three-nucleon systems and saturation\textemdash that is, the occurrence of a minimum of the energy per particle $E(\varrho)/A$  at non-vanishing density $\varrho_0 = 0.16 \ {\rm fm}^{-3}$\textemdash in SNM.

%%%%%%%%%%%%%%%%%%%%%%%%%%%%%%%%%%%%%%%
Phenomenological
models of the NNN force such as the UIX potential~\cite{UIX_2,UIX_1} are written in the form
\begin{align}
\label{NNN:split}
V_{ijk}=V_{ijk}^{2\pi}+V_{ijk}^{R} \ ,
\end{align}
where $V_{ijk}^{2\pi}$ is the attractive Fujita-Miyazawa potential~\cite{Fujita}, 
describing two-pion-exchange processes associated with the appearance of a $\Delta$ resonance
in the intermediate state, while $V_{ijk}^{R}$ is a purely phenomenological repulsive term.
The UIX model involves two parameters, determined from fits to the empirical data  
performed using the AV18+UIX nuclear Hamiltonian.
The strengths of the two-pion exchange contribution and the isoscalar repulsive term are fixed to simultaneously reproduce the observed ground-state energies of \isotope[3][]{He} and \isotope[4][]{He}, as well as the saturation density of SNM, $\rho_0 = 0.16$ fm$^{-3}$.
%%%%%%%%%%%%%%%%%%%%%%%%%%%%%%%%%%%%%%%

The non relativistic Hamiltonian employed to obtain the effective interaction discussed in this 
work comprises the Argonne $v_6^\prime$ 
(AV6P) NN potential~\cite{V6P}, determined {\it projecting} the full AV18 potential onto the basis of the six operators appearing in the expression of the one-pion-exchange potential, defined by Eqs.\eqref{av18:2}
and \eqref{S12}, and the UIX NNN potential.
AV6P predicts the binding energy and electric quadrupole moment of the deuteron with accuracy of 1\%, and 4\%, respectively, and provides an excellent fit of the elastic NN scattering phase shifts in the $^1{\rm S}_0$ 
channel\footnote[3]{We will adopt the spectroscopic notation, according to which the two-nucleon
states labelled ${^1}{\rm S}_0$ and ${^1}{\rm P}_1$ correspond to spin $S=0$ and orbital angular momentum 
$\ell = 0$ and 1, respectively.}\textemdash which is dominant in pure neutron matter\textemdash up to laboratory  energies $\sim600$ MeV, well above the pion-production threshold.

%%%%%%%%%%%%%%%%%%%%%%%%%%%%%%%%%%%%%%%%%%%%%
\begin{figure}[hbt]
    \centering
    \includegraphics[scale=0.65]{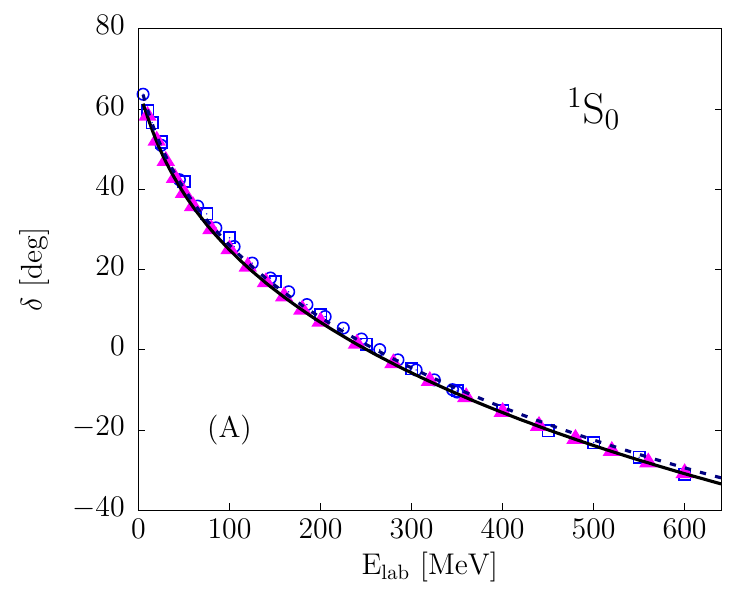}
    \includegraphics[scale=0.65]{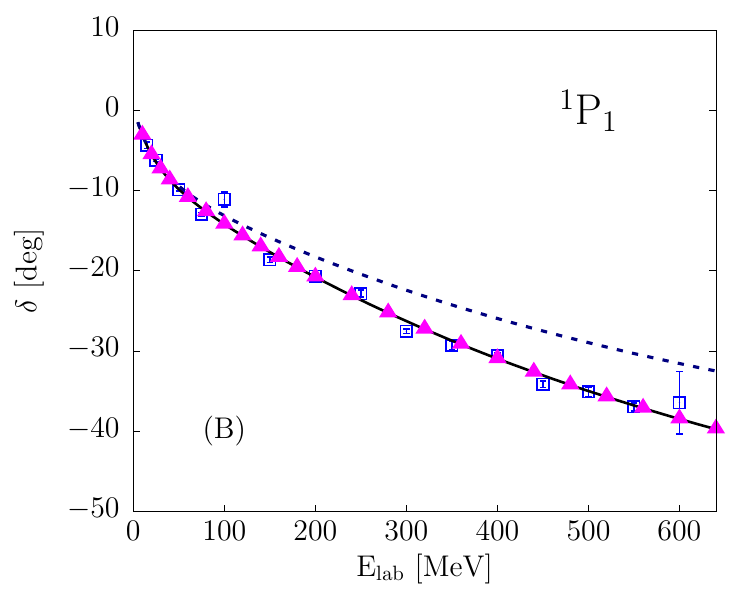}
    \caption{Proton-neutron scattering phase shifts in the $^1{\rm S}_0$ (A) and $^1{\rm P}_1$ (B)
channels, plotted as a function of the kinetic energy of the beam particle in the laboratory frame. The results obtained using the AV6P
and the truncated AV18 potentials are represented by solid and dashed lines, respectively. The triangles correspond to the full AV18 potential. The data are taken from Refs.~\cite{Nijmegen1,Nijmegen2} (circles) and~\cite{ SAID} (squares). }
   \label{phase_shift}
\end{figure}
%%%%%%%%%%%%%%%%%%%%%%%%%%%%%%%%%%%%%%%%%%%%%

It must be pointed out that the procedure employed to obtain the AV6P potential effectively takes into account interactions described by components
of the AV18 potential having $p>6$. This feature is clearly illustrated in Fig.~\ref{phase_shift}, providing a comparison between the neutron-proton scattering phase shifts in
the $^1{\rm S}_0$ and $^1{\rm P}_1$ channels
computed using the AV6P potential and those obtained from the AV18 potential neglecting the
 terms $v^{p>6}$. It appears that
the AV6P model provides a description of the data comparable to that obtained from the full AV18 potential in both channels.
On the other hand, the truncated
AV18 potential, while being capable to predict the $^1{\rm S}_0$ phase shifts, conspicuously fails to reproduce the data in the $^1{\rm P}_1$ channel at energies
larger than $\sim$100 MeV.
This pattern suggests that the contributions of angular momentum-dependent interactions, which turn out
to be significant, are, in fact, largely effectively included in the projected AV6P potential.

The results of a study performed using accurate many-body
techniques also show that the EOSs of PNM computed using the AV6P potential and the full AV18 potential remarkably
agree with one another. At saturation density, the discrepancy between the energies per particle obtained from 
auxiliary field diffusion Monte Carlo (AFDMC) calculations using the two potential
models turns out to be $\sim$10-15\%~\cite{Piarulli_benchmark}.

%%%%%%%%%%%%%%%%%%%%%%%%%%%%%%%%%%%%%%%%%%%%%%%%%%%%%%%
\subsection{Relativistic corrections}
\label{relacorr}
%%%%%%%%%%%%%%%%%%%%%%%%%%%%%%%%%%%%%%%%%%%%%%%%%%%%%%%

The relativistic description of nuclear systems is a long-standing challenge, which has been confronted within different theoretical approaches. Here, we outline the formalism originally developed by  Krajcik and Foldy~\cite{Foldy-Krajcik}, and further elaborated by Friar~\cite{Friar75}. 

The authors of Ref.~\cite{Foldy-Krajcik} derived the relativistic Hamiltonian 
of a system with fixed number of particles from a covariant approach based on the well established conceptual framework of non-relativistic quantum mechanics.
In this context, {\it relativistic covariance} refers to the 
requirement that the descriptions of physical phenomena observed in different inertial frames can be obtained from one another by performing a Poincar\'e transformation.
Such a requirement entails the definition of a representation of the Poincar\'e group in the Hilbert space 
associated with the observed system, and the derivation of the relativistic Hamiltonian from the commutation rules of the Poincar\'e algebra. A clear and exhaustive discussion of the assumptions underlying this approach  can be found in Ref.~\cite{Foldy1}.

The relativistic Hamiltonian of an interacting many-particle system, $H_{\rm R}$, is defined as the sum of the relativistic single-particle kinetic energies, the 
interaction potentials and the associated relativistic boost corrections. The explicit expression comprising interactions involving up to three particles can be written as  
\begin{align}
H_{\rm R}=\sum_{i=1}^A \sqrt{m_i^2+\vb{p}_i^2} & +\sum_{j>i=1}^A\left[\tilde{v}_{ij} +\delta v_{ij}(\vb{P}_{ij})\right] \\
\nonumber &+\sum_{k>j>i=1}^A \left[\widetilde{V}_{ijk}+\delta V_{ijk}(\vb{P}_{ijk})\right] \ ,
\end{align}
to be compared to Eq.~\eqref{H:A}.
In the above equation, $\tilde{v}_{ij}$ and $\widetilde{V}_{ijk}$ denote the NN and NNN potentials in the rest frames of the interacting nucleons, defined by the conditions
\begin{equation}
\vb{P}_{ij}=\vb{p}_i+\vb{p}_j=0 \ , 
\end{equation}
and
\begin{equation}
\vb{P}_{ijk}=\vb{p}_i+\vb{p}_j+\vb{p}_k=0 \ , 
\end{equation}
with ${\bf p}_i$, ${\bf p}_j$ and ${\bf p}_k$ being the nucleon momenta.
The boost corrections $\delta v_{ij}$ and $\delta V_{ijk}$ originate from the motion of the two- and 
three-nucleon center of mass in the rest frame of the $A$-nucleon system. They  satisfy
the obvious requirements
\begin{equation}
	\delta v_{ij}(\vb{P}_{ij}=0) = \delta V_{ijk}(\vb{P}_{ijk}=0) = 0 \ .
\end{equation}

Because, in principle, $\tilde{v}_{ij}$ and the non relativistic ${v}_{ij}$ of Eq.~\eqref{H:A} can be both obtained from a fit to experimental data~\cite{CPS}, 
one can reasonably assume that relativistic effects taken into account using $\tilde{v}_{ij}$ in conjunction with  square root kinetic energies are, in fact, implicitly 
included in ${v}_{ij}$. 

In the 1970s, Krajcik and Foldy~\cite{Foldy-Krajcik} and Friar~\cite{Friar75} obtained an expression of the boost correction $\delta v({\bf P})$\footnote[3]{The subscripts specifying the labels of the interacting particles will be omitted whenever not necessary.} from an expansion in powers of ${\bf P}^2/4 m^2$. 
On the other hand, relativistic corrections to the NNN potential, whose contribution to the energy is known to be much smaller than the one arising from NN interactions, 
have been generally neglected~\cite{boost,CPS}.
The present discussion will be limited to the calculation of the leading order contribution to $\delta v({\bf P})$  within the scheme  
of Forest {\it et al.}~\cite{boost}. This approximation appears to be amply justified by the results of numerical calculations, showing that 
the probability of ${\bf P}^2$ being larger than $m^2$ is, in fact, quite small; see, e.g., Ref.~\cite{CPS}.

The starting point is the decomposition 
\begin{equation}
H=H_0+H_I \ , 
\label{expansion:H}
\end{equation}
where $H_0$ is the Hamiltonian describing two non interacting nucleons and $H_I$,  containing all interaction 
terms, can be conveniently written in the form
\begin{equation}
H_I=v+\delta v + ...\ .
\end{equation}
The potential $v$, independent of $\vb{P}$, can be assumed to be of order $1/m$, because in nuclear systems interaction energy and non-relativistic 
kinetic energy are known to be of comparable magnitude. The correction $\delta v $ is of order $1/m^3$, and the ellipsis represents 
terms of order $1/m^4$ or higher.
The presence of $H_I$ entails the appearance of a corresponding term in the boost generator, that can be written in the form
\begin{equation}
\vb{K}=\vb{K}_0+\vb{K}_I \ .
\label{expansion:K}
\end{equation}
The leading contributions to $H_I$ and $\vb{K}_I$ have been derived in Refs.~\cite{Foldy-Krajcik,Friar75}. 
Substitution of the resulting expressions into the relevant commutation relations of the Poincar\'e algebra yields 
the boost interaction at order $\vb{P}^2/4m^2$~\cite{boost}
\begin{align}
\label{deltav}
\delta v(\vb{P}) =& -\frac{\vb{P}^2}{8m^2}\tilde{v}+\frac{1}{8m^2}[(\vb{P}\cdot\vb{r})(\vb{P}\cdot\grad),\tilde{v}] \\
\nonumber
&+\frac{1}{8m^2}[(\boldsymbol{\sigma}_1-\boldsymbol{\sigma}_2)\times(\vb{P}\cdot\grad),\tilde{v}] \ ,
\end{align}
where the potential in the NN center-of-mass frame, $\tilde{v}$, includes all relativistic corrections independent of $\vb{P}$, and the gradient operator
acts on the relative coordinate.
The first two terms in the right hand side of the above equation are associated with the relativistic energy-momentum 
relation and Lorentz contraction, respectively, whereas the last contribution\textemdash which has been shown to be negligibly small 
in light nuclei~\cite{CPS}\textemdash  
account for Thomas precession and quantum mechanical effects. 

The results reported in this work have been obtained using the Hamiltonian 
\begin{align}
\nonumber
H^*_{\rm NR} = \sum_{i=1}^A  \frac{{\bf p}_i^2}{2m} & + \sum_{j>i=1}^A  [ v_{ij} + \delta v({\vb P}_{ij}) ] \\
\label{hstar}
 & + \sum_{k>j>i=1}^A  V^*_{ijk} \ ,
\end{align}
in which $V^*_{ijk}$ denotes the NNN potential modified to account  for the presence of the boost interaction $\delta v$.
The calculation of  $\delta v$ has been carried out  following the approximations employed in Ref.~\cite{APR}, 
which amounts to neglecting the second line of Eq.~\eqref{deltav} as well as   
non-static NN interactions, corresponding to the $p>6$ components of the NN potential.
%%%%%%%%%%%%%%%%%%%%%%%%%%%%%%%%%%%%%%%%%%%%%%%%%%%%%%%
\section{Effective Hamiltonian}
\label{renormalisation}
%%%%%%%%%%%%%%%%%%%%%%%%%%%%%%%%%%%%%%%%%%%%%%%%%%%%%%%
Within the approach developed by the authors of Ref.~\cite{BL:2017}\textemdash a simplified  implementation of which was originally proposed in 
Ref.~\cite{Cowell:2002bv}\textemdash  the effective interaction is defined by the equation 
\begin{align}\label{exp-ham}
\langle\Psi_0|H_{\text{NR}}|\Psi_0\rangle & =T_{F}+\langle\Phi_0|V_{\rm eff}|\Phi_0\rangle 
\end{align}
where $H_{\text{NR}}$ is  the non relativistic Hamiltonian of Eq.~\eqref{H:A}, $T_F$ denotes the energy of the non interacting system, and 
\begin{align}
\label{veff:2}
V_{\rm eff} = \sum_{j>i=1}^A v^{\tn{eff}}_{ij} \ , 
\end{align}
with the effective NN potential $v^{\tn{eff}}_{ij}$ written as in Eq.\eqref{eq:NN_1} and including $p\leq 6$ contributions only. 

The nuclear matter ground state  $|\Psi_0\rangle$ is 
obtained from the Fermi gas ground state, $|\Phi_0\rangle$, through the transformation 
\begin{equation}\label{eq:corr_state}
	|\Psi_0\rangle=\frac{\mathcal{F}|\Phi_0\rangle}{\sqrt{\langle\Phi_0|\mathcal{F}^{\dagger} \mathcal{F}|\Phi_0\rangle}} \ , 
\end{equation}
with
\begin{equation}
\label{def:F}
\mathcal{F}=\mathcal{S}\prod_{j>i}F_{ij} \ .
\end{equation}
The two-nucleon correlation functions appearing in the right-hand side of the above equation are meant to 
embody the majority of dynamical effects. As a consequence, they can be conveniently written in the form 
\begin{align}
F_{ij}=\sum_{p=1}^6  f^{p}(r_{ij}) O^{p}_{ij} \ , 
\label{eq:F_1}
\end{align}
reminiscent of the NN potential of Eq.~\eqref{eq:NN_1}. Note that because, in general, $[F_{ij},F_{jk}] \neq 0$, the product in Eq.\eqref{def:F} 
needs to be symmetrised through the action of the operator $\mathcal{S}$.
The most prominent effect of the inclusion of correlations is the appearance of {\it screening}, that is, a strong suppression of the probability 
to find two nucleons within the range of the non perturbative repulsive core of the NN interaction. 

Using the formalism described in Ref.~\cite{BL:2017}, the effective Hamiltonian of Eq.~\eqref{exp-ham} can be readily 
used to obtain the energy of nuclear matter with arbitrary neutron excess. In the case one-component Fermi liquids, 
such as PNM and SNM, the ground state expectation value of the effective interaction reduces to the form\footnote[3]{Hereafter, $\langle \ldots \rangle$
will denote an expectation value in the Fermi gas ground state}
\begin{equation}\label{eq:pot_bare_2b}
\langle~V_{\rm eff}~\rangle= A~\frac{\varrho}{2}\int d^3r_{12}\,\mbox{CTr}\left[ v^{\tn{eff}}_{12}\left(1-\widehat{P}^{\sigma\tau}_{12}\ell^2(k_Fr_{12})\right)\right] \ , 
\end{equation}
where $k_F$ is the Fermi momentum and $r =| \vb{r}_1-\vb{r}_2 | $. In the above equation, $\text{CTr}~[\ldots ]$ denotes the
trace operation  in spin-isospin space, normalised such that $\text{CTr}~[\, \mathbb{1} \,] = 1$, and $\widehat{P}^{\sigma\tau}_{12}$ is the spin-isospin exchange operator, 
defined as
\begin{equation}\label{eq:pair_exchange}
    \widehat{P}^{\sigma\tau}_{12}=\frac{1}{4}(1+ \boldsymbol{\sigma}_1\cdot\boldsymbol{\sigma}_2)(1+\boldsymbol{\tau}_1\cdot\boldsymbol{\tau}_2)  \ .
\end{equation}
The function $\ell(k_Fr)$\textemdash often referred to as Slater function\textemdash is trivially related to the density matrix of the 
non interacting system at density $\varrho$ through
\begin{align}
\nonumber
   \ell(k_Fr) & = \frac{1}{A} \sum_{\bf k}  \theta(k_F - |{\bf k}|)~e^{i {\bf k} \cdot {\bf r} } \\ 
   & = 3 \left[ \frac{\sin{(k_Fr)}- (k_Fr)  \cos{(k_Fr)}}{(k_Fr)^3}\right] \ ,
\end{align}
with  $\theta(x)$ being the Heaviside step function.

The effective interaction is obtained from Eq.~\eqref{exp-ham} using the cluster expansion formalism~\cite{CLARK197989}, which amounts to rewrite
the left-hand side in the form
\begin{equation}
\label{cluster:expansion}
    \langle\Psi_0|{H}_{\text{NR}}|\Psi_0\rangle=T_{F}+(\Delta E)_2+ \ldots \ ,
\end{equation}
where $(\Delta E)_n$ is the contribution arising from subsystems\textemdash or clusters\textemdash comprising $n$ particles.

Keeping only two-body terms, one obtains the simple expression 
\begin{equation}\label{eq:eff}
	v^{\tn{eff}}_{ij}=  -\frac{1}{m} ( \boldsymbol{\nabla}F_{ij} )^2 + F_{ij}v_{ij}F_{ij}  \ , 
\end{equation}
which clearly illustrates how the the bare NN potential is {\it renormalised} by the action of the correlation functions.  

%%%%%%%%%%%%%%%%%%%%%%%%%%%%%%%%%%%%%%%%%%%%%%%%%%%%%%%
\begin{figure}[t!]
    \centering
    \includegraphics[scale=0.55]{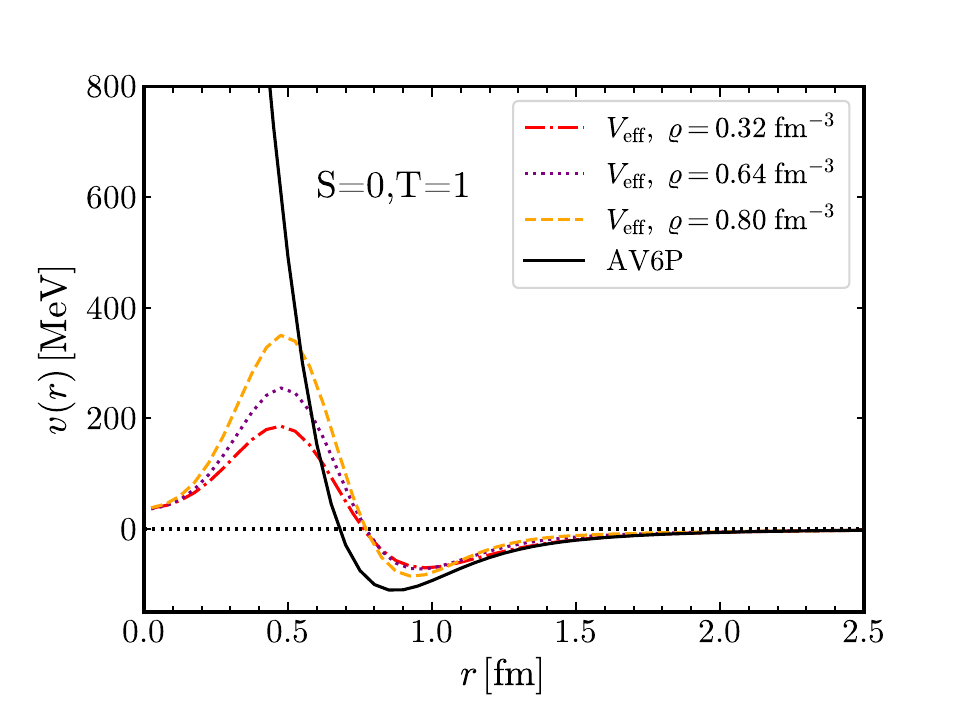}
    \caption{Radial and density dependence of the CBF effective interaction obtained from the AV6P+UIX Hamiltonian. For comparison, the thick solid
    line shows the behaviour of the bare AV6P potential.}
    \label{renorm}
\end{figure}
%%%%%%%%%%%%%%%%%%%%%%%%%%%%%%%%%%%%%%%%%%%%%%%%%%%%%%%

The shape of the $f^p(r)$ is found by solving the set of Euler-Lagrange equations obtained from the minimisation of the ground-state energy,  
evaluated at the two-body cluster level using the NN potential
\begin{equation}
\label{eq:quenched}
	{\bar v}_{ij}=\sum_p [\delta_{p1} + (1- \delta_{p1}) \alpha] v^p(r_{ij})O^p_{ij} \,
\end{equation}
with $\alpha$ being a parameter that accounts for the quenching of NN interactions in the nuclear medium.
The correlation functions satisfy the boundary conditions
\begin{equation}
\label{parameters:1}
    	f^{p}(r\ge d^p)=\delta_{1p} \ \ , \ \ 
    	\left.\frac{df^p(r)}{dr}\right|_{{d^p}}=0 \ , 
\end{equation}
where, following Ref.~\cite{LAGARIS1981331}, we have set 
\begin{align}
\label{parameters:2}
d^{p<5}=d_c \ \ \ \ , \ \ \ \ d^{p=5,6}=d_t \ .
\end{align}

From the above discussion, it follows that, for any density $\varrho$, the correlation functions associated with a given nuclear Hamiltonian depend on three
parameters: the quenching factor $\alpha$ and the two relaxation distances $d_c$ and $d_t$.
In principle, they should be regarded as variational parameters, to be determined from the minimisation of the ground-state energy within  
an advanced computational method, such as the FHNC/SOC summation scheme or the variational Monte Carlo technique. According to the philosophy 
underlying the effective interaction formalism, on the other hand, they are adjusted in such a way as to reproduce a set of  
{\it target} results\textemdash
obtained from accurate non perturbative calculations\textemdash using the potential $V_{\rm eff}$ defined by Eqs.~\eqref{exp-ham}\textendash \eqref{veff:2} and first order perturbation theory.
This procedure has been successfully implemented to derive an effective interaction from the AV6P+UIX Hamiltonian, with the  role of
target results being played by the EOSs of SNM and PNM, obtained starting from the same Hamiltonian using the 
FHNC/SOC and the AFDMC computational techniques, respectively.
%of Akmal {\it et al.}~\cite{Akmal:1997, APR} corresponding to the non relativistic AV18+UIX Hamiltonian.  
%These results, obtained without inclusion of the relativistic boost corrections, will be referred to as APR1 EOSs.

As pointed out previously, the determination of $V_{\rm eff}$ from Eq.~\eqref{exp-ham} involves the calculation of the expectation value of the
the nuclear Hamiltonian in the correlated ground sate, performed using the cluster expansion method. The lowest order approximation
allowing to take into account both the NN and NNN potentials involves two- and three-body cluster 
contributions. The resulting expression can be cast in the form
\begin{align}
\label{eq:eff_int_newdef}
    \langle~V_{\text{eff}}~\rangle  & = \langle~\mathcal{F}^{\dagger}[T,\mathcal{F}]~\rangle|_{\text{2b}}
     +\langle~\mathcal{F}^{\dagger} V_{\text{NN}} \mathcal{F}~\rangle|_{\text{2b}}  \\
     \nonumber
    &  +  \langle~\mathcal{F}^{\dagger} V_{\text{NNN}} \mathcal{F}~\rangle|_{\text{3b}} \ , 
\end{align}
where $T$ denotes the $A$-body kinetic energy operator, 
\begin{align}
\label{define:V}
V_{\text{NN}} = \sum_{j>i=1}^{A} v_{ij} \ \  ,  \ \ V_{\text{NNN}} = \sum_{k>j>i=1}^A V_{ijk} \ , 
\end{align}
and the subscripts $2b$ and $3b$ label two- and three-nucleon cluster 
contributions, respectively.

It should be emphasised that the correlation operator $\mathcal{F}$ appearing in Eq.~\eqref{eq:eff_int_newdef} is {\it not} 
the same as the one determined performing a minimisation of the ground-state energy of nuclear matter. The values of the parameters $\alpha$, $d_c$, and $d_t$ 
entering the calculation of  the right-hand side of Eq.~\eqref{eq:eff_int_newdef} are instead adjusted to satisfy Eq.~\eqref{exp-ham} using the effective interaction $V_{\rm eff}$\textemdash obtained from the AV6P+UIX Hamiltonian\textemdash and the target results appearing in the the left-hand side. 
%and the APR1 EOSs as target results in the left-hand side. 
The radial and density dependence of CBF effective potential obtained from the AV6P+UIX Hamiltonian are illustrated in Fig.~\ref{renorm}

%%%%%%%%%%%%%%%%%%%%%%%%%%%%%%%%%%%%%%%%%%%%%%%%%%%%%%%
\section{Boost corrections to the nuclear matter EOS}
\label{results}
%%%%%%%%%%%%%%%%%%%%%%%%%%%%%%%%%%%%%%%%%%%%%%%%%%%%%%%

As a first approximation, the impact of including relativistic boost corrections into the effective interaction of Ref.~\cite{BL:2017} has been investigated by treating the corresponding contribution to the ground-state energy as a perturbation. This procedure involves fitting the correlation operator $\mathcal{F}$ so that the effective interaction reproduces the non-relativistic EOS of PNM and SNM. The resulting expression reads
\begin{equation}
\label{eq:pert_boost}
    \widetilde{V}_{\text{eff}}=V_{\text{eff}}+\langle~\mathcal{F}^{\dagger}~ \delta V~\mathcal{F}~\rangle|_{2b} = V_{\text{eff}} + \delta V_{\text{eff}}\ , 
\end{equation}
with
\begin{align}
\label{def:deltav}
%\delta v =  \sum_{j>i} \delta v_{ij}   , 
\delta V =  \sum_{j>i} \delta v_{ij}   , 
\end{align}
where, following Ref.~\cite{APR}, the form of $\delta v_{ij}$ is approximated by
\begin{align}
\label{approx:deltav}
\delta v_{ij} = & -\frac{\vb{P}_{ij}^2}{8m^2}{v_{ij}}+
\frac{1}{8m^2}[(\vb{P}_{ij}\cdot\vb{r})(\vb{P}_{ij}\cdot\grad),{v_{ij}}] \ .
\end{align}
%%%%%%%%%%%%%%%%%%%%%%%

%%%%%%%%%%%%%%%%%%%%%%%
In the above equation, the gradient operates on the relative coordinate and $v_{ij}$ is a NN potential whose operator structure is described by the $O^{p \leq 6}$ of Eqs.~\eqref{av18:2} and \eqref{S12}.
%%%%%%%%%%%%%%%%%%%%%%%%%%%%%%%%%%%%%%%%%%%%%%%%%%%%%%%
\subsection{AV18 and AV6P Hamiltonians}
\label{results:NN}
%%%%%%%%%%%%%%%%%%%%%%%%%%%%%%%%%%%%%%%%%%%%%%%%%%%%%%%
To single-out the effects of $\delta v$, we first considered Hamiltonian models that include the NN potential alone.

The EOSs of PNM and SNM computed by the authors of Refs.~\cite{Akmal:1997,APR} using the AV18 and AV18+$\delta v$ Hamiltonians have been compared to those obtained from the corresponding effective interactions, derived from the AV6P potential using the AV18 EOSs as target.

\begin{figure}
    \centering
    \includegraphics[scale=0.60]{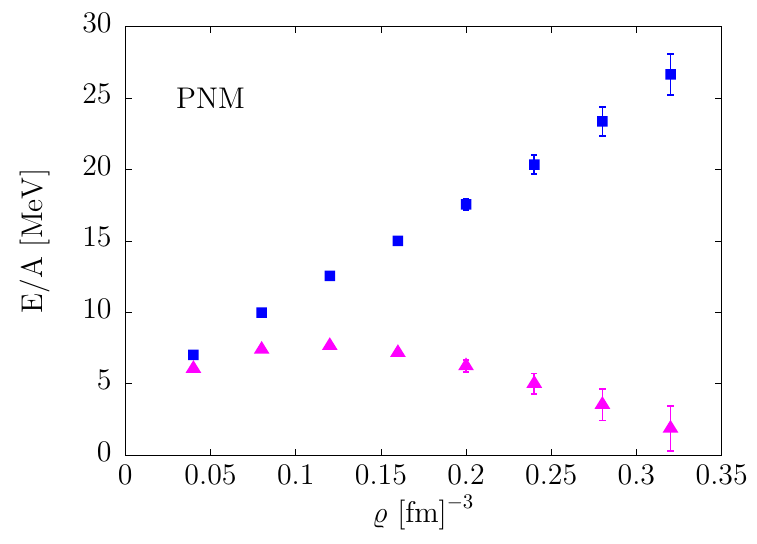}
    \caption{Energy per particle of  PNM, obtained from AFDMC calculations performed 
    including NN interactions only. Squares and triangles correspond to the 
   AV6P potential and to the truncated AV18 potential discussed in the text.}
   \label{fig:compare}
\end{figure}

It must be kept in mind that, while having the same operator structure, the truncated AV18 potential\textemdash employed to evaluate the contribution of the relativistic boost interaction 
in Ref.~\cite{APR}\textemdash  differs significantly from the projected AV6P potential. The discrepancies observed in the NN scattering phase shifts, discussed in  Section~\ref{nuclear:hamiltonian}, also strikingly emerge in the density dependence of the energy per nucleon of PNM computed using the AFDMC; see Fig.~\ref{fig:compare}. 

%%%%%%%%%%%%%%%%%%%%%%%%
%%%%%%%%%%%%%%%%%%%%%%%%%

\begin{figure}[t!]
    \centering
    \includegraphics[scale=0.55]{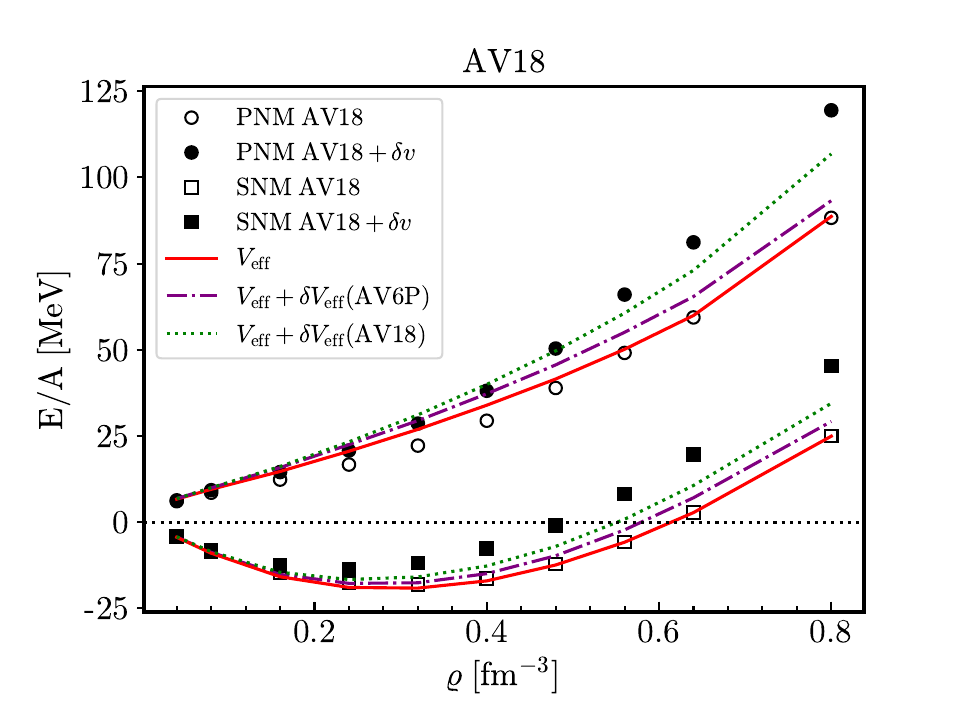}
    \caption{Energy dependence of the binding energy per particle of PNM and SNM obtained using Hamiltonians including only NN potentials.The meaning of the curves is explained in the text.}
    \label{fig:AV18_1}
\end{figure}

Figure~\ref{fig:AV18_1} shows a comparison between the variational ground-state energies of PNM and SNM determined  by Akmal {\it et al.} using the AV18 and AV18+$\delta v$ Hamiltonians and the corresponding results obtained using the CBF effective interaction with and without inclusion of the relativistic correction defined by
Eq.~\eqref{eq:pert_boost}. The calculations carried out using the AV6P 
and the truncated AV18 potential, labelled $\delta V_{\rm eff}(AV6P)$ and $\delta V_{\rm eff}(AV18)$, respectively, yield significantly different energies over the whole density range, with $\delta V_{\rm eff}(AV18)$ being appreciably larger than $\delta V_{\rm eff}(AV6P)$ and closer to Akmal's results. It is apparent that both the $V_{\rm eff} + \delta V_{\rm eff}(AV6P)$ and $V_{\rm eff} + \delta V_{\rm eff}(AV18)$ models fail to reproduce the results obtained using the AV18+$\delta v$ Hamiltonian.

\begin{figure}[!h]
    \centering
    \includegraphics[scale=0.55]{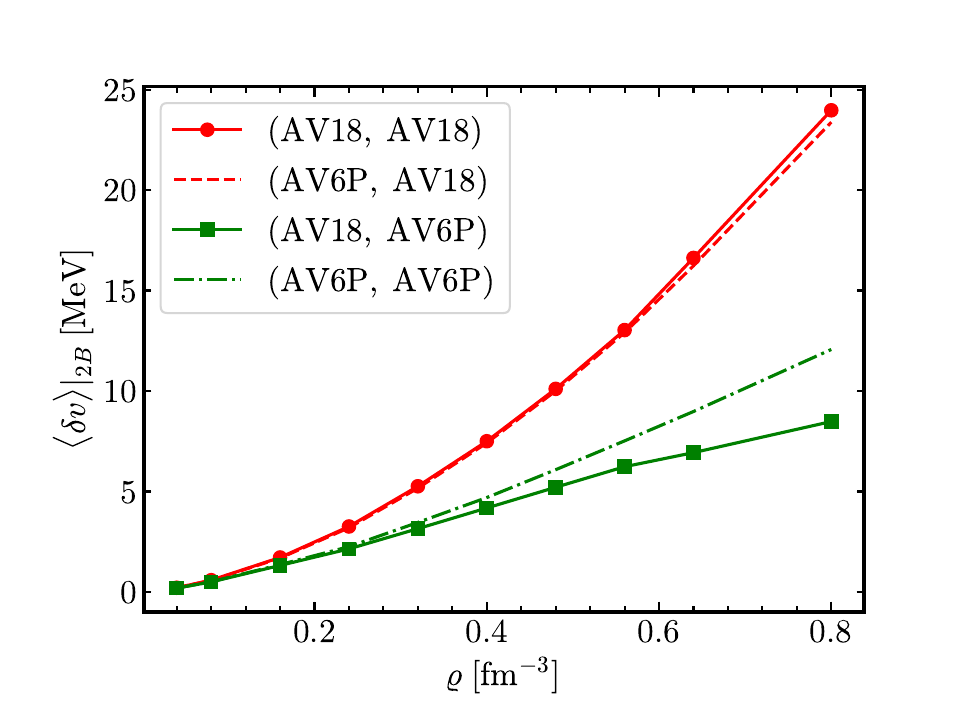}
    \caption{Density dependence of the relativistic boost correction $\delta V_{\rm eff}$\textemdash defined by  Eq.~\eqref{eq:pert_boost}\textemdash in PNM. The results have been obtained using different NN potentials and correlation functions. The meaning of the curves is explained in the text. 
    }
   \label{fig:5}
\end{figure}

To investigate the origin of the pattern seen in Fig.~\ref{fig:AV18_1}, we study the sensitivity of the boost interaction energy to modifications of the NN correlation functions $f^p(r)$. In Fig.~\ref{fig:5}, the curves labelled (X,Y)\textemdash with X,Y~$\equiv$ AV18 or AV6P\textemdash represent results obtained using the potential model X for the solution of the Euler-Lagrange equations determining the shape of the $f^p(r)$, and model Y in the calculation of the boost interaction energy. It appears that the observed behaviour is primarily driven by the potential used in the calculation of the boost interaction energy, although for X$\equiv$AV6P some differences between the energies corresponding to different correlation functions emerge at high densities. 
%%%%%%%%%%%%%%%%%%%%%%%%%%%%%%%%%%%%%%%%%%%
%%%%%%%%%%%%%%%%%%%%%%%%%%%%%%%%%%%%%%%%%%%%

%%%%%%%%%%%%%%%%%%%%%%%%%%%%%%%%%%%%%%%%%%%%%%%%%%%%%%%
%%%%%%%%%%%%%%%%%%%%%%%%%%%%%%%%%%%%%%%%%%%%%%%%%%%%%%%

%%%%%%%%%%%%%%%%%%%%%%%%%%%%%%%%%%%%%%%%%%%%%%%%%%%%%%%
%%%%%%%%%%%%%%%%%%%%%%%%%%%%%%%%%%%%%%%%%%%%%%%%%%%%%%%

%%%%%%%%%%%%%%%%%%%%%%%%%%%%%%%%%%%%%%%%%%%%%%%%%%%%%%%
\subsection{AV18+UIX and AV6P+UIX Hamiltonians}
\label{results:NNN}
%%%%%%%%%%%%%%%%%%%%%%%%%%%%%%%%%%%%%%%%%%%%%%%%%%%%%%%
The formalism described in the previous sections has been generalised to allow the treatment of models of nuclear dynamics including a NNN potential, the corresponding effective interaction being defined by Eq.\eqref{eq:eff_int_newdef}. In this Section, we will compare our findings to the EOSs of PNM and SNM  reported by Akmal {\it et al.} in Ref.~\cite{APR}. For convenience, the EOSs obtained with the the AV18+UIX and AV18+$\delta v$+UIX$^*$ Hamiltonians will be referred to as APR1 and APR2, respectively. We recall the reader that 
the presence of the repulsive boost interaction brings about a significant modification of the UIX potential. The strength of the isoscalar repulsive contribution
to the resulting NNN interaction, denoted UIX$^*$, turns out to be reduced by a factor $\gamma = 0.63$~\cite{Forest_Arriaga,APR}.

Since we are working in perturbation theory, the parameters $\alpha$, $d_c$ and $d_t$ entering the definition of the two-nucleon correlation functions, discussed in Section~\ref{renormalisation}, have been adjusted to reproduce the APR1 EOS. The energies per nucleon corresponding to the APR2 model reported in Ref.~\cite{APR} have been obtained by adding to the corresponding APR1 results the contribution of the boost interaction and the associated modification of the NNN potential, computed in first order perturbation theory.
The resulting expression can be written in the form
\begin{align}
\label{enter:gamma}
 \left( \frac{E}{A} \right)_{\text{APR2}} &= \left( \frac{E}{A} \right)_{\text{APR1}}\nonumber \\ 
 &+\frac{1}{A} \langle \mathcal{F}^{\dagger}[ \delta V + (\gamma-1) V^R ] \mathcal{F}  \rangle,
\end{align}
with the correlation operator $\mathcal{F}$ being determined from minimisation of the expectation value of the AV18+UIX Hamiltonian in the correlated ground state, and
\begin{align}
V^R = \sum_{k>j>i} V^R_{ijk} \ . 
\end{align}

To identify the effects of relativistic corrections, we have evaluated the contribution of the effective potential to the ground-state energy using different 
prescriptions. In the first scenario, relativistic boost corrections are neglected altogether, whereas 
the two additional definitions are
\begin{align}
\delta{V_{\rm eff}} = \langle {\mathcal F}^\dagger [ \delta V + (\gamma-1) V^R ] \mathcal F \rangle \ , 
 \label{prescr:1}
\end{align}
and 
\begin{align}
\delta{V^*_{\rm eff}} = (\gamma - 1) \langle {\mathcal F}^\dagger V^R \mathcal F \rangle \ .
 \label{prescr:2}
\end{align}

In all three cases the correlation functions determining the effective interaction are the same, with the corresponding parameters being adjusted using as target results the APR1 EOSs of Ref.~\cite{APR}.

\begin{figure}
    \centering
    \includegraphics[scale=0.575]{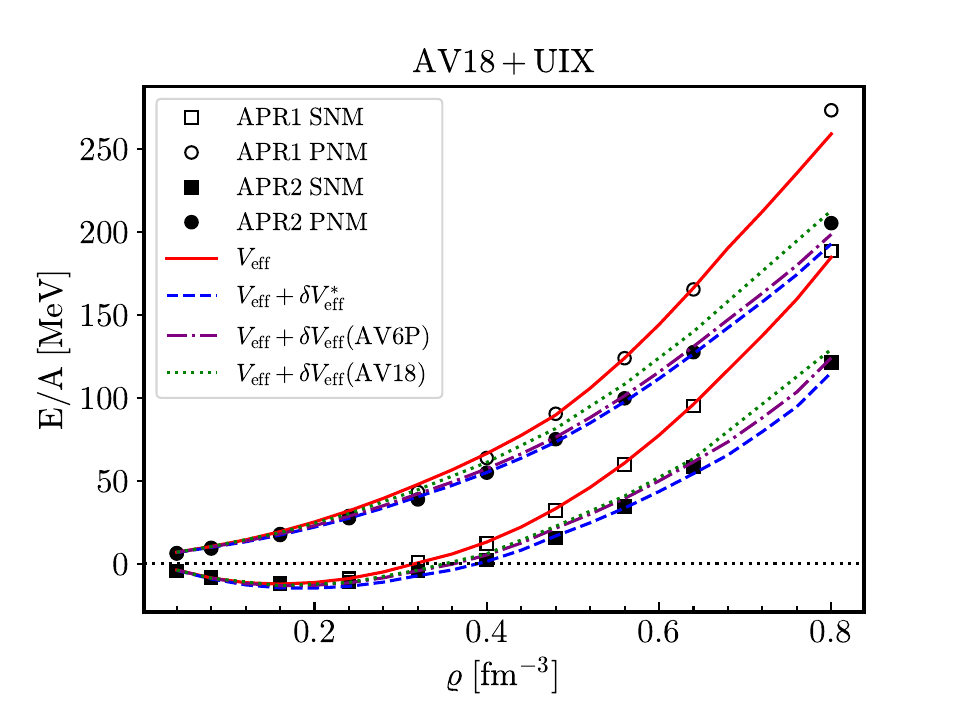}
    \caption{Energy dependence of the binding energy per particle of PNM and SNM obtained using Hamiltonians including both NN and NNN potentials.
The meaning of the curves is explained in the text.
    }
    \label{fig:3}
\end{figure}

%Figure~\ref{fig:3} shows the results obtained with these models. 
The solid lines of Fig.~\ref{fig:3}\textemdash which reproduce the APR1
EOSs by construction\textemdash represent the results obtained using $V_{\rm eff}$ defined by Eq.~\eqref{eq:eff_int_newdef}, 
while the dashed lines correspond to calculations carried out using $\delta V^*_{\text{eff}}$ of Eq.~\eqref{prescr:2}.
The results obtained including the full relativistic correction of Eq.\eqref{prescr:1} and using the AV6P and the
truncated AV18 in the calculation of $\delta V$ are displayed by the dot-dash and dotted lines, respectively.

It clearly appears that the differences originating from the use of the AV6P or the truncated AV18 potential in 
the calculation of the boost interaction energy, while being still visible, are largely obscured by the size of the 
associated modification of the repulsive NNN potential. 

%%%%%%%%%%%%%%%%%%%%%%%%%%%%%%%%%%%%%%%%%%%%%%%%%%%%%%%
%%%%%%%%%%%%%%%%%%%%%%%%%%%%%%%%%%%%%%%%%%%%%%%%%%%%%%%

%%%%%%%%%%%%%%%%%%%%%%%%%%%%%%%%%%%%%%%%%%%%%%%%%%%%%%%%%%%%%%%%%%%%%%%%%%%%%%%%%%%%%%%%%%%%%%%%%%%%%%%
\section{Boost correction to the CBF Effective Interaction}
\label{boost:veff}
%%%%%%%%%%%%%%%%%%%%%%%%%%%%%%%%%%%%%%%%%%%%%%%%%%%%%%%%%%%%%%%%%%%%%%%%%%%%%%%%%%%%%%%%%%%%%%%%%%%%%%%

In the previous sections, the relativistic boost interaction has been treated in perturbation theory, 
which amounted to adding to the CBF effective potential of Eq.~\eqref{eq:eff_int_newdef} the expectation value of $\delta V$ in the correlated ground state, defined by Eq.~\eqref{eq:pert_boost}. Here, we will discuss a procedure to take into account the boost interaction at operator level, which implies determining the relativistic effective potential, denoted $V^\prime_{\rm eff}$, from 
the equation 
\begin{equation}
\label{def:Veffprime}
\langle {\mathcal F}^\dagger {\rm H}^\prime{\mathcal F} \rangle =T_F
+\langle {V}_{\rm eff}^\prime \rangle,
\end{equation}
where
\begin{equation}
\label{Hprime}
{\rm H}^\prime = \sum_i\frac{\vb{p}^2_i}{2m}+\sum_{j>i}(v_{ij}+\delta v_{ij})+\sum_{k>j>i}V^*_{ijk} \ ,
\end{equation}
the operator structure of $V^\prime_{\rm eff}$ being the same as that of 
Eq.\eqref{eq:NN_1} with $p\leq 6$. 

It clearly appears that the procedure based on Eqs.~\eqref{def:Veffprime} and \eqref{Hprime} 
allows to take into account the effects of boost interactions in the determination of the 
radial dependence of the correlation functions $f^p(r)$, which in turn drives the behaviour 
of the effective interaction.

\begin{figure*}[htb]
    	\centering
     \quad
    	\subfigure[]{\includegraphics[scale=0.5]{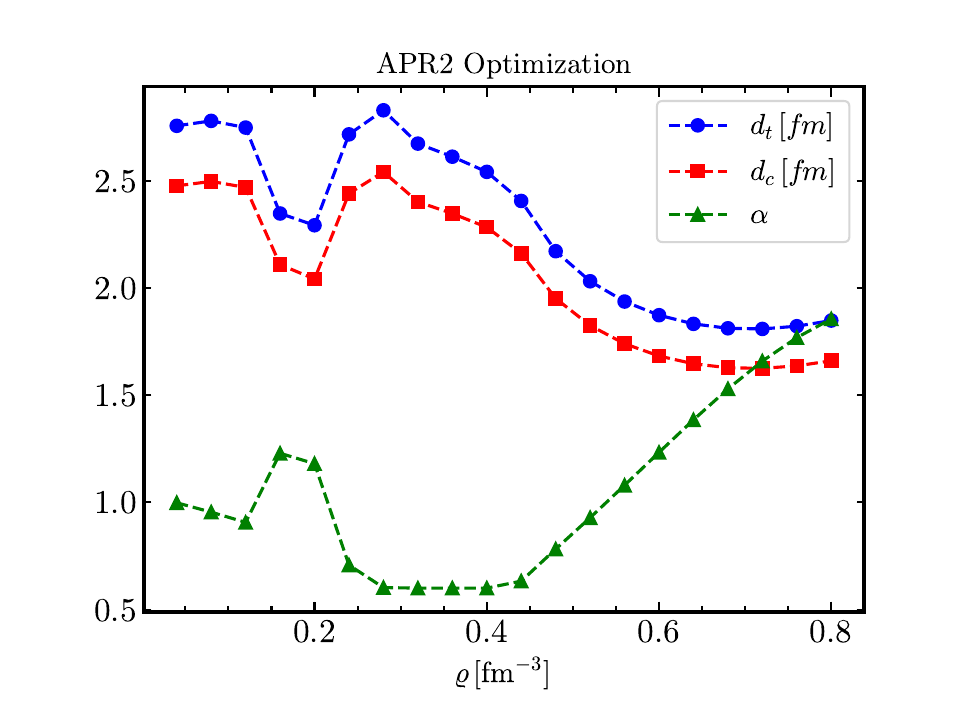}}\qquad
    	\subfigure[]{\includegraphics[scale=0.5]{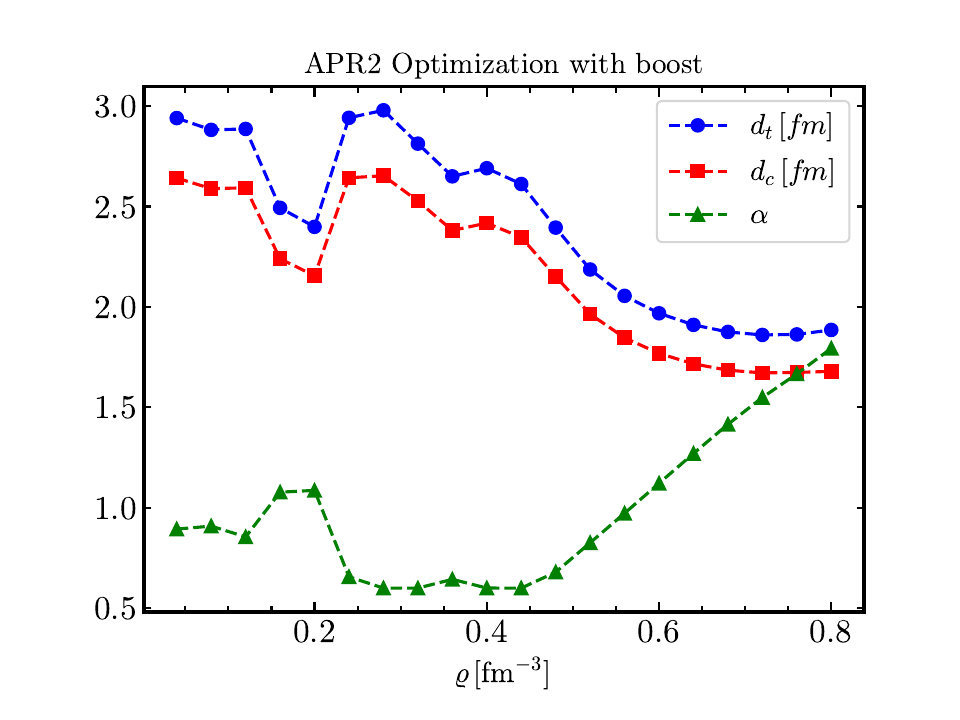}}\\
    	\subfigure[]{\includegraphics[scale=0.5]{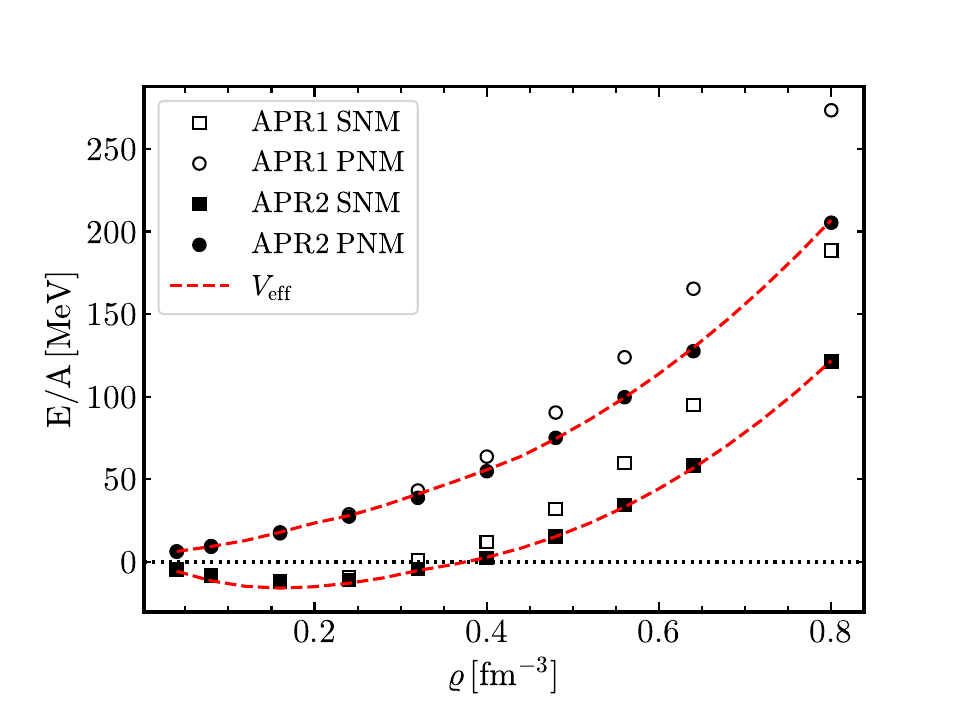}}\quad
    	\subfigure[]{\includegraphics[scale=0.5]{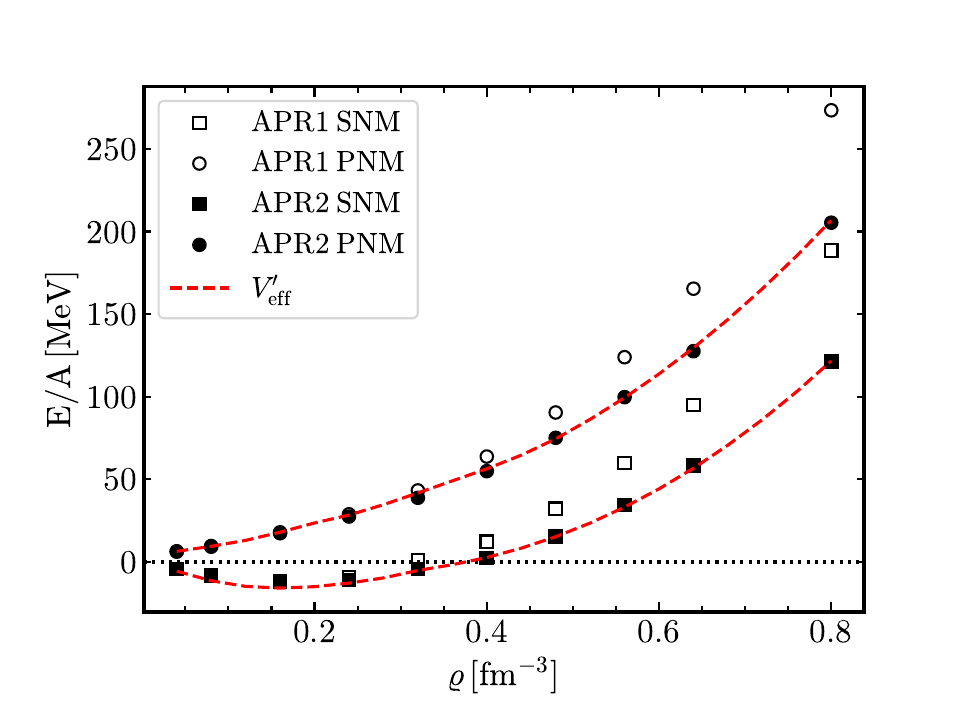}}
    	\caption{The top panels show the correlation ranges and quenching parameters, defined by 
     Eqs.~\eqref{eq:quenched}-\eqref{parameters:2} obtained optimizing the effective interaction to reproduce APR2 EOSs without (a) and with (b) the inclusion of  boost corrections in its definition. 
     The corresponding energies per nucleon of both SNM and PNM are displayed in panels (c) and (d). 
     Squares and circles represent the results of Akmal {\it et al.}~\cite{APR}.}
    	\label{fig:APR2boost_opti_UIX}
	\end{figure*}

The numerical results obtained from the above method are summarised in Fig.~\ref{fig:APR2boost_opti_UIX}.
In analogy to Sect.~\ref{results:NNN}
we have considered two different definitions of the effective potential. The first one, which takes into 
account boost corrections only through the replacement $V_{\rm NNN}~\to~V^*_{\rm NNN}$, 
is written in the form
\begin{align}
\label{eq:eff2_final_noboost}
\langle~V_{\text{eff}}~\rangle  & = \langle~\mathcal{F}^{\dagger}[T,\mathcal{F}]~\rangle|_{\text{2b}}
     +\langle~\mathcal{F}^{\dagger} V_{\text{NN}} \mathcal{F}~\rangle|_{\text{2b}}  \\
     \nonumber
    &  +  \langle~\mathcal{F}^{\dagger} V^*_{\text{NNN}} \mathcal{F}~\rangle|_{\text{3b}} \ , 
    \end{align}
    reminiscent of the non relativistic expression of Eq.~\eqref{eq:eff_int_newdef}.
The second prescription, on the other hand, is obtained from Eq.~\eqref{def:Veffprime}, and 
includes relativistic corrections to the NN potential and the quenching of the NNN interaction. The corresponding expression reads
%$\delta v$ of Eq.~\eqref{def:deltav} 
\begin{align}
    \label{eq:eff2_final_boost}
    \langle~V^\prime_{\text{eff}}~\rangle  & = \langle~\mathcal{F}^{\dagger}[T,\mathcal{F}]~\rangle|_{\text{2b}}
     +\langle~\mathcal{F}^{\dagger} ( V_{\text{NN}} + \delta V )\mathcal{F}~\rangle|_{\text{2b}}  \\
     \nonumber
    &  +  \langle~\mathcal{F}^{\dagger} V^*_{\text{NNN}} \mathcal{F}~\rangle|_{\text{3b}} \ .
\end{align}

The effective potentials defined by Eqs.~\eqref{eq:eff2_final_noboost} and~\eqref{eq:eff2_final_boost} have 
 been both determined in such a way as to reproduce the APR2 EOSs, obtained using the AV18+$\delta v$+UIX$^*$
 Hamiltonian. The details of the calculations can be found in the appendix.

The density dependence of the parameters $\alpha$, $d_c$ and $d_t$ obtained by fitting 
the target EOSs with the effective potentials of Eqs.~\eqref{eq:eff2_final_noboost} and~\eqref{eq:eff2_final_boost} are displayed in panels (a) 
and (b) of Fig.~\ref{fig:APR2boost_opti_UIX}, respectively, while panels (c) and (d) show the 
corresponding energies per particle of PNM and SNM. The jump in the values of the parameters
determining the shape of the correlation functions, clearly visible at $\varrho\approx0.2\,{\rm fm}^{-3}$, 
is likely to reflect the appearance of the spin-isospin ordered phase\textemdash associated with the 
occurrence of neutral pion condensation\textemdash discussed in Ref.\cite{APR}. The small kink at  
$\varrho\approx0.32\,{\rm fm}^{-3}$ may also be ascribed to this phase transition in SNM.
It appears that the inclusion of boost interactions has only a marginal impact on the parameter values, and the target energies can be accurately reproduced with both choices of the effective potential.  
 
%We indeed recall that we are optimizing the free parameters of the effective interaction in order to simultaneously reproduce both SNM and PNM energies.

%%%%%%%%%%%%%%%%%%%%%%%%%%%%%%%%%%%%%%%%%%%%%%%%%%%%%%%%%%%%%%%%%%%%%%%%%%%%%%%%%%%%%%%%%%%%%%%%%%%%%%%%%%%%%%%%
%%%%%%%%%%%%%%%%%%%%%%%%%%%%%%%%%%%%%%%%%%%%%%%%%%%%%%%%%%%%%%%%%%%%%%%%%%%%%%%%%%%%%%%%%%%%%%%%%%%%%%%%%%%%%%%%

\begin{figure*}[!htb]
    	\centering
    	\subfigure[]{\includegraphics[scale=0.5]{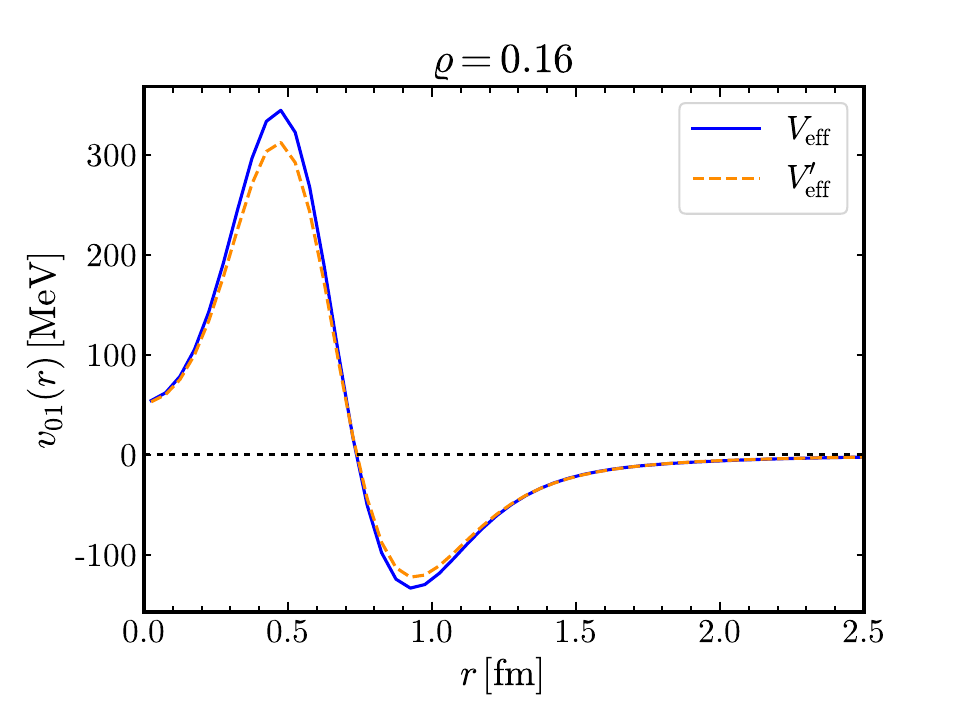}}
    	%\subfigure[]{\includegraphics[scale=0.5]{figures_effUIX/effboost_032.pdf}}\\
    	\subfigure[]{\includegraphics[scale=0.5]{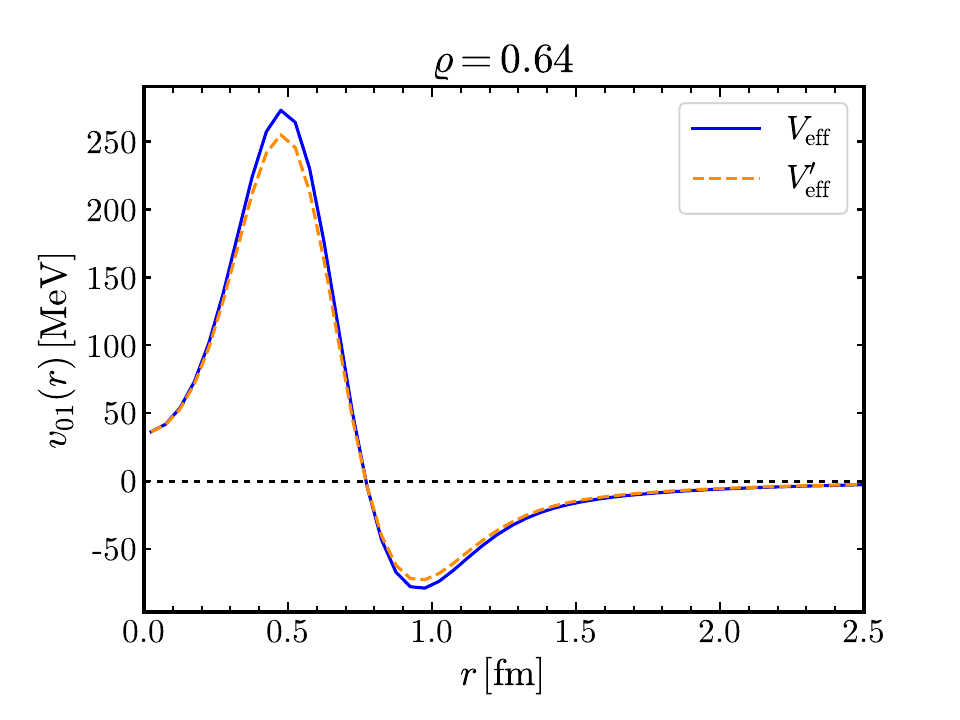}}
    	%\subfigure[]{\includegraphics[scale=0.5]{figures_effUIX/effboost_080.pdf}}
    	\caption{ Radial dependence of the CBF effective interaction in the $S=0$, $T=1$ channel 
     at density $\varrho = 0.16$ (a) and $0.64$ fm$^{-3}$ (b). The solid line represents  
     the potential $V_{\rm eff}$ obtained from Eq.~\eqref{eq:eff2_final_noboost} using the UIX$^*$ 
     NNN potential, while the dashed line corresponds to the effective interaction 
     $V^\prime_{\rm eff}$, determined including the contribution of relativistic boost interactions according to 
     Eq.~\eqref{eq:eff2_final_boost}.
     }
    	\label{fig:poteff_radial2_UIX}
	\end{figure*}

Figure~\ref{fig:poteff_radial2_UIX} illustrates the radial behaviour of the CBF effective potential  
acting in the two-nucleon channel of spin-isospin $S=0$ and $T=1$ at density $\varrho = 0.16$ and
$0.64$ fm$^{-3}$. The solid line  
corresponds to $V_{\rm eff}$ of Eq.~\eqref{eq:eff_int_newdef}, while the dashed line represents the effective interaction 
     $V^\prime_{\rm eff}$ determined including relativistic boost interactions according to 
     Eq.~\eqref{eq:eff2_final_boost}. The UIX$^*$ NNN potential has been used in both cases. Note that, because the effective interactions $V_{\rm eff}$ and $V^\prime_{\rm eff}$ are optimised to reproduce the same APR2 EOSs, the values of the 
     adjustable parameters $\alpha$, $d_c$ and $d_t$ determining the shape of the correlation functions involved in their definitions are not the same. The modification arising from boost interactions
     are clearly visible. 
     
%%%%%%%%%%%%%%%%%%%%%%%%%%%%%%%%%%%%%%%%%%%%%%%%%%%%%%%%%%%%%%%%%%%%%%%%%%%%%%%%%%%%%%
\section{Summary and outlook}
\label{outlook}
%%%%%%%%%%%%%%%%%%%%%%%%%%%%%%%%%%%%%%%%%%%%%%%%%%%%%%%%%%%%%%%%%%%%%%%%%%%%%%%%%%%%%%

We have investigated the impact of including relativistic boost corrections into 
the density-dependent effective interaction of Ref.~\cite{BL:2017}, derived from the AV6P+UIX model of  
the nuclear Hamiltonian using the CBF formalism and the cluster expansion technique. 
The contribution of the boost interaction $\delta V$ has been considered both at average level, by
treating its expectation value in the correlated ground state as a perturbation, 
and at microscopic level, by explicitly taking into account  $\delta V$ in the determination the effective potential.

The form of $\delta V$ has been obtained following the procedure described in  Refs.~\cite{Forest_Arriaga,boost}, designed to be used with static NN potentials.
We found significant differences between the contributions of the boost interaction to 
nuclear matter energy computed with either the AV6P potential or the truncated version of the AV18 model employed by Akmal {\it et al.}~\cite{APR}, in which all angular-momentum-dependent 
terms are neglected. These discrepancies suggest that the projection method
used by the authors of Ref.~\cite{V6P} to obtain the AV6P potential allows to effectively capture some
important features of non static interactions.
The contribution of the boost interaction computed using the truncated AV18 interaction turns out to 
be considerably larger than that obtained from the AV6P at all densities. 

The inclusion of a repulsive correction to the NN potential obviously entails a corresponding 
decrease of the repulsive component of the NNN potential, $V^R$, needed to preserve the capability
of the Hamiltonian to explain the measured binding energies of \isotope[3][]{He} and \isotope[4][]{He} and the empirical value of the saturation 
density of SNM. Our analysis shows that the resulting $\sim30$\% reduction of the strength 
of repulsive NNN interactions is, in fact, the dominant relativistic  correction, leading to 
a significant softening of the nuclear matter EOS at high densities.

In this context, it is worth mentioning that additional constraints on NNN interactions in 
dense matter may be provided by multimessenger neutron star observations. The results of the 
analyses performed by the authors of Refs.\cite{Maselli:2020uol,Sabatucci2022} suggest that the
existing and forthcoming data have the potential to allow an accurate determination of 
the parameter $\gamma$ of Eq.~\eqref{enter:gamma}, providing a measure of the relativistic 
boost correction to the strength of $V^R$.

The analysis described in this article is meant to be a first exploratory step towards a 
fully consistent inclusion of relativistic boost corrections into the effective interaction
of Ref.~\cite{BL:2017}. The achievement of this goal\textemdash which involves using 
correlation functions optimised to reproduce the EOSs of PNM and SNM obtained from the AV6P+UIX$^*$
Hamiltonian\textemdash will extend the density region in which the CBF effective Hamiltonian can be 
expected to provide a reliable description of nuclear matter properties relevant to neutron star
structure and dynamics. 

Relativistic modifications of the three-nucleon potential have been also recently
discussed by the authors of Ref.~\cite{LucaG}, who derived the boost 
corrections to the leading order NNN contact interaction from the constraints imposed by Poincar\'e algebra on non relativistic dynamics. However, their impact on nuclear properties has not been studied yet.

As a final remark, it should be emphasised that, in assessing the role of relativistic corrections to the dynamics of dense matter, one has to keep in mind that the approach followed in our study is only applicable in the density regime in which nuclear matter can be described in terms of nucleon degrees of freedom. 
The data recently reported by the Neutron Star Internal Composition 
Explorer (NICER)\cite{Riley:2021pdl,Riley:2019yda}, 
showing that neutron stars of $\sim1.4$ and $\sim2.1$ solar masses have similar radii, provide convincing evidence that the limit of applicability of this description may be as high as $\sim4 \varrho_0$.   

\acknowledgments

This research was funded by the U.S. Department of Energy, Office of Science, Office of Nuclear Physics, under contract DE-AC02-06CH11357 (A.L.), the NUCLEI SciDAC program (A.L.), and the Italian National Institute for Nuclear Research (INFN), under grant TEONGRAV (O.B. and A.S.).

%%%%%%%%%%%%%%%%%%%%%%%%%%%%%%%%%%%%%%%%%%%%%%%%%%%%%%%%%%%%%%%%%%%%%%%%%%%%%%%%%%%%%%
\appendix
\label{appendix:A}
\begin{widetext}
%%%%%%%%%%%%%%%%%%%%%%%%%%%%%%%%%%%%%%%%%%%%%%%%%%%%%%%%%%%%%%%%%%%%%%%%%%%%%%%%%%%%%%
\section{Boost corrections to the CBF effective interaction}\label{chapter:Boost}
%%%%%%%%%%%%%%%%%%%%%%%%%%%%%%%%%%%%%%%%%%%%%%%%%%%%%%%%%%%%%%%%%%%%%%%%%%%%%%%%%%%%%%

In this Appendix, we provide details on the calculation of the effective interaction %$V^\prime_{\text{eff}}$ 
defined by Eq.~\eqref{eq:eff2_final_boost}, in which the corrections arising from relativistic boost interactions 
are explicitly taken into account at operator level. In view of the fact that the contribution of 
$\delta V$ is computed in the two-body cluster approximation, we can limit our discussion to the case of a Hamiltonian 
comprising two-nucleon terms only, which can be written  
\begin{equation}\label{eq:hboost}
	H = T + V_{\rm NN} + \delta V = \sum_i\frac{\vb{p}_i^2}{2m}+\sum_{j>i}\left(v_{ij}+\delta v_{ij}\right).
\end{equation}
The effective interaction is obtained by expanding the expectation value of $H$ in the correlated
ground state, and retaining only the two-body cluster contribution $(\Delta E)_{2}$; see Eq.~\eqref{cluster:expansion}. 
In this case, the boost correction, which will be denoted $\delta v^{\rm eff}_{ij}$, is implicitly defined by the equation 
\begin{align}
\label{deltaE2}
(\Delta E)_{2} = \langle~{\mathcal F}^\dagger [T,\mathcal{F}]~\rangle|_{2b} +  \langle~{\mathcal F}^\dagger \sum_{j>i} ( v_{ij} + \delta  v_{ij} ) \mathcal{F}~\rangle|_{2b}
 = \langle~\sum_{j>i} ( v^{\rm eff}_{ij} + \delta  v^{\rm eff}_{ij} )~\rangle \ , 
\end{align}
where 
\begin{equation}\label{eq:eff}
	v^{\tn{eff}}_{ij}=  -\frac{1}{m} ( \boldsymbol{\nabla}F_{ij} )^2 + F_{ij}v_{ij}F_{ij}  \ , 
\end{equation}
with $F_{ij}$ being the NN correlation function of Eq.~\eqref{eq:F_1}. Because both $v_{ij}$ and $F_{ij}$ depend linearly on the operators $O^{p\leq6}$  defined by Eqs.~\eqref{av18:2} and \eqref{S12}, one can exploit the algebra
\begin{equation}\label{AV6_algebra}
O^p_{ij}O^q_{ij}=\sum_{r=1}^{6} K^{pqr}O^r_{ij} \ ,
\end{equation}
with the coefficients $K^{pqr}$ given in, e.g., Ref.\cite{bob_vijay_rmp}, to cast the NN effective potential in the form
\begin{equation}
%	v_{ij}^{\tn{eff}}=\sum_pv^{p}_{\tn{eff}}(\varrho,r_{ij})O_{ij}^p
v_{ij}^{\tn{eff}}=\sum_{p=1}^6v^{{\rm eff},p}(r_{ij})O_{ij}^p \ .
\end{equation}
%{\color{magenta}
%If the operator structure of  $\delta v^{\rm eff}_{ij}$ can be cast in the same form as that of $v_{ij}^{\tn{eff}}$, given by the above equation, its expression 
%can be obtained from the equation  
%\begin{equation}\label{eq:boost_effective_exp}
%	\langle~\mathcal{F}^\dagger \sum_{j>i}\delta v_{ij} \mathcal{F}~\rangle|_{2b}=\frac{\varrho}{2}\int d^3 r_{12}\,\mbox{CTr}\left[\sum_p 
%	\delta v^{\tn{eff},p}(r_{12})O^p_{12}\left(1-\widehat{P}^{\sigma\tau}_{12}\ell^2(k_Fr_{12})\right)\right].
%\end{equation}
%}
%which can be obtained by inserting into the expression of $\langle\mathcal{F}^\dagger \delta v_{12}|\mathcal{F} \rangle|_{2b}$ the identity
%\begin{equation}
%   \left[1-\widehat{P}^{\sigma\tau}_{12}\ell^2(k_Fr)\right]^{-1}\left[1-\widehat{P}^{\sigma\tau}_{12}\ell^2(k_Fr)\right] \equiv \mathbb{1}\ .
%\end{equation}

The form of the boost interaction employed in this work, given by~Eq.\eqref{approx:deltav}, has been derived using the approximations discussed in 
Refs.~\cite{boost,APR}. It does not include the terms arising from Thomas Precession and the commutators involving the spin-dependent 
part of the NN potential present in Eq.~\eqref{deltav}, whose contributions to the binding energies of \isotope[3][]{He} and \isotope[4][]{He} have been found to be 
negligible~\cite{forest_thomas}. For a static NN potential written as in Eq.~\eqref{eq:NN_1} one obtains 
\begin{equation}
\begin{split}
\delta v_{12} =\sum_{p=1}^6 \left[-\frac{\vb{P}^2}{8m^2}v^p(r_{12})+\frac{(\vb{P}\cdot {\hat{\bf r}}_{12} )^2}{8m^2}r_{12}\frac{dv^p}{dr_{12}}\right]O^p_{12}+ \sum_{p=1}^6 \frac{(\vb{P}\cdot\vb{r}_{12})}{8m^2}v^p(r_{12})(\vb{P}\cdot\grad) O^p_{12} \ ,
\end{split}
\end{equation}
where ${\hat{\bf r}}_{12} = {\bf r}_{12}/r_{12}$.
Note that the last term in the right-hand side of the above equation involves the gradient of the tensor operator $S_{12}$,  whose expectation value vanishes in  
unpolarised matter. For this reason, here we limit ourselves to consider the expression
\begin{equation}
\delta v_{12}=\sum_{p=1}^6\left[-\frac{\vb{P}^2}{8m^2}v^p(r_{12})+\frac{(\vb{P}\cdot {\hat{\bf r}}_{12})^2}{8m^2}r_{12}\frac{dv^p}{dr_{12}}\right]O^p_{12}
= \sum_{p=1}^6 \delta v^p(r_{12}) O^p_{12} \ .
\end{equation}

To obtain the expression of the relativistic boost correction  we need to calculate the two-body cluster contribution to the
expectation value of  $\delta V$ in the correlated ground state.
As a first step, we consider
the expectation value in the Fermi gas ground state, which can be readily evaluated and split into direct and exchange contributions 
according to 
\begin{align}
\label{FG:1}
\frac{1}{A}~\langle \delta V \rangle = \frac{1}{A}~\langle \sum_{j>i} \delta v_{ij} \rangle = \frac{1}{A}~\langle \delta V \rangle_D  - 
\frac{1}{A}~\langle \delta V \rangle_E  \  , 
\end{align} 
where
\begin{equation}
\begin{split}
\frac{1}{A}~\langle\delta V \rangle_D=\frac{1}{5}\frac{\varrho}{8m^2}k_{F}^2\sum_p\int d^3r_{12} \left[r_{12}\frac{dv^p(r_{12})}{dr_{12}}-3v^p(r_{12})\right]\mbox{CTr}[O^p_{12}] \ ,
\end{split}
\end{equation}
 and
\begin{equation}
\begin{split}
\frac{1}{A} \langle\delta V \rangle_E=\frac{\varrho}{16m^2}\sum_p\int d^3r_{12} \left[r_{12}\frac{dv^p(r_{12})}{dr_{12}}I_2(k_F,r_{12})-v^p(r)I_1(k_F,r_{12})\right]\mbox{CTr}[O^p\widehat{P}^{\sigma\tau}_{12}] \ ,
\end{split}
\end{equation}
with
\begin{align}
I_1(k_F,r)&=2[\ell^\prime(k_{F}r)]^2-2\ell(k_{F}r)\grad^2_{r}\ell(k_{F}r)\ , \nonumber\\
I_2(k_F,r)&=2[\ell^\prime(k_{F}r)]^2-2\ell(k_{F}r)\ell''(k_{F}r)\nonumber \ .
\end{align}
Here a prime indicates derivative with respect to $r$, 
\begin{equation}
\grad^2_r\equiv\frac{1}{r}\frac{\partial^2}{\partial r ^2}r \ , 
\end{equation}
 and $k_F$ denotes the Fermi momentum. Note that in this work we only consider one-component Fermi liquids, such as~PNM and SNM.

A more compact expression can be obtained collecting the direct and exchange terms in the right-hand side of Eq.~\eqref{FG:1}. The result can be 
written in the form
\begin{equation}
	\frac{1}{A}\langle~\delta V \rangle=\frac{\varrho}{2} \sum_{p=1}^6 \left\{\int d^3r_{12}\,A^p(k_F,r_{12})\mbox{CTr}[O_{12}^p]-\int d^3r_{12}\,B^p(k_F,r_{12})\mbox{CTr}[O_{12}^p\widehat{P}^{\sigma\tau}_{12}]\right\}.
\end{equation}
where
\begin{equation}
\begin{split}
	A^p(k_F,r)&\equiv\frac{1}{8m^2}\frac{2}{5}k_F^2\left[r \frac{dv^p(r)}{dr}-3v^p(r)\right] \ ,\\
	B^p(k_F,r)&\equiv\frac{1}{8m^2}\left[r \frac{dv^p(r)}{dr}I_2(k_F,r)-v^p(r)I_1(k_F,r)\right] \ .
\end{split}
\end{equation}
Finally, because the spin-isospin exchange operator can be written in terms of the operators of Eq.~\eqref{av18:2} as
\begin{equation}
	\widehat{P}^{\sigma\tau}_{12}=\frac{1}{4}\sum_{q=1}^{4}O_{12}^q,
\end{equation}
we can use the algebra \eqref{AV6_algebra} to obtain
\begin{equation}
\frac{1}{A}\langle \delta V\rangle=\frac{\varrho}{2}  \sum_{p,m=1}^6 \int d^3r_{12}\,\left[ A^p(k_F,r_{12})\delta_{pm}-\frac{1}{4}\sum_{q=1}^{4}K^{pmq}B^p(k_F,r_{12})\right] \mbox{CTr}[O_{12}^m].
\end{equation}

Let us now consider the two-body cluster approximation to the expectation value of $\delta V$ in the correlated ground state, derived in, e.g., Ref.~\cite{CLARK197989}, whose expression reads
\begin{align}
\label{derive:delta}
\frac{1}{A}~\biggl\langle~\mathcal{F}^\dagger \biggl(\sum_{j>i}\delta v_{ij}\biggr)~\mathcal{F}~\biggr\rangle\Biggr|_{2b}=\frac{A-1}{2}~\biggl\langle~\sum_{qmn}~f^q(r_{12}) \delta v^m(r_{12})f^n(r_{12})O^q_{12}O^m_{12}O^n_{12}~\biggr\rangle
\end{align}
It clearly appears that by rewriting  $\delta v^p$ in the form
\begin{align}
\label{newdelta:2}
 \delta v^p(r_{12})  =  -\frac{\vb{P}^2}{8m^2} w^p(r_{12}) +\frac{(\vb{P} \cdot {\hat{\bf r}_{12}} )^2}{8m^2} g^p(r_{12})
\end{align}
with
\begin{equation}
\label{def:wg}
w^p(r_{12}) =  v^p(r_{12}) \ \ \ \ \ \ , \ \ \ \ \ \ g^p(r_{12}) = r_{12}\frac{dv^p}{dr_{12}} \ , 
\end{equation}
and exploiting again the algebra of Eq.~\eqref{AV6_algebra}, the right-hand side of Eq.~\eqref{derive:delta}  can by simply 
computed replacing
\begin{equation}
\begin{split}
w^p(r)&\rightarrow\sum_{mn\ell q}K^{mn\ell}K^{q\ell p}f^q(r)\,v^m(r)\,f^n(r) \ ,\\g^p(r)&\rightarrow\sum_{mn\ell q}K^{mn\ell}K^{q\ell p}f^q(r)\,r\frac{dv^m}{dr}\,f^n(r) \ , 
\end{split}
\end{equation}
in the expectation value of $\delta V$ in the Fermi gas ground state.

From now on, summation over repeated indices of the $O^{p\leq 6}$ operators of Eq.~\eqref{AV6_algebra} will be understood, as well as the subscripts labeling the interacting particles. 
By defining the quantities
\begin{equation}
	\begin{split}
    	\xi^m(k_F,r)&=\left\{A^m(k_F,r)-\frac{1}{4}\sum_{q=1}^{4}K^{pqm}B^p(k_F,r)\right\} ,\\
    	M_{pm}&=\frac{1}{4}\sum_{q=1}^{4}K^{pqm}.
	\end{split}
\end{equation}
we can rewrite the Fermi gas result in the form
\begin{align}
  \frac{1}{A}\langle \delta V \rangle&=\frac{\varrho}{2}\int d^3r\,\mbox{CTr}[O^m]\xi^m(k_F,r)\\
  &=\frac{\varrho}{2}\int d^3r\,\mbox{CTr}[O^m]\left[\delta_{mn}-\ell^2(k_Fr)M_{mn}\right]\nonumber\left[\delta_{np}-\ell^2(k_Fr)M_{np}\right]^{-1}\xi^p(k_F,r) \ .
\end{align}
Since
\begin{equation}\label{eq:inverse}
   \left[\delta_{mp}-\ell^2(k_Fr)M_{mp}\right]^{-1}= \frac{1}{1-\ell^4(k_Fr)} \left[\delta_{mp}+\ell^2(k_Fr)M_{mp}\right],
\end{equation}
we finally obtain
\begin{equation}\label{eq:final_boost_exp}
	\frac{1}{A}\langle\delta V\rangle=\frac{\varrho}{2}\int d^3r\,\mbox{CTr}[O^m]\left[\delta_{mp}-l^2(k_Fr)M_{mp}\right]\delta v^p(k_F,r)
\end{equation}
where we have defined
\begin{equation}
	\delta v^p(k_F,r)\equiv\frac{1}{1-l^4(k_Fr)} \left[\delta_{pm}+l^2(k_Fr)M_{pm}\right] \xi^m(k_F,r).
\end{equation}
By recalling that the operator $M$ is linked to the spin-isospin exchange operator $\widehat{P}^{\sigma\tau}$, Eq.~\eqref{eq:final_boost_exp} can be written as
\begin{equation}
	\frac{1}{A}\langle\delta V\rangle=\frac{\varrho}{2}\int d^3r\,\mbox{CTr}\left[O^p\left(1-l^2(k_Fr)\widehat{P}^{\sigma\tau}\right)\right]\delta v^p(k_F,r)
\end{equation}
which is the expression we were looking for.

Finally we are going to derive the expression of Eq.~\eqref{eq:inverse}. Since the operator we want to invert has the form
\begin{equation}
	\mathbb{1}-l^2M
\end{equation}
we have that
\begin{eqnarray}
  \left(\mathbb{1}-l^2M\right)\left(\mathbb{1}+l^2M^{-1}\right)&=&(1-l^4)\mathbb{1}-l^2M+l^2M^{-1}.
\end{eqnarray}
Therefore if we show that
\begin{equation}
	M=M^{-1}
\end{equation}
we have done. We start from the identity
\begin{equation}
	O^mO^pO^q=K^{mpl}K^{lqs}O^s
\end{equation}
and summing over $p,q=1,...,4$ we have
\begin{eqnarray}
	\sum_{p,q=1}^4O^mO^pO^q=O^m\sum_{p=1}^4O^p\sum_{q=1}^4O^q=O^m(4\widehat{P}^{\sigma\tau})(4\widehat{P}^{\sigma\tau})=16\,O^m(\widehat{P}^{\sigma\tau})^2=16\,O^m.
\end{eqnarray}
Therefore we can write
\begin{eqnarray}
O^m=\frac{1}{16}\sum_{p,q=1}^4O^mO^pO^q=\left(\frac{1}{4}\sum_{p=1}^4K^{mpl}\right)\left(\frac{1}{4}\sum_{q=1}^4K^{lqs}\right)O^s=M_{ml}M_{ls}O^s
\end{eqnarray}
which entails
\begin{equation}
	M_{ml}M_{ls}=\delta_{ms}\;\Rightarrow \;M=M^{-1}.
\end{equation}

\end{widetext}

\bibliographystyle{utphys}
\bibliography{Ref}

\providecommand{\href}[2]{#2}\begingroup\raggedright\begin{thebibliography}{10}

\bibitem{bob}
R.~Wiringa, ``From deuterons to neutron stars: variations in nuclear many-body theory,'' \href{http://dx.doi.org/10.1103/RevModPhys.65.231}{{\em Rev. Mod. Phys.} {\bfseries 65} (Jan, 1993) 231--242}. \url{https://link.aps.org/doi/10.1103/RevModPhys.65.231}.

\bibitem{Coester:NM}
F.~Coester, S.~Pieper, and F.~Serduke, ``Relativistic effects in phenomenological nucleon-nucleon potentials and nuclear matter,'' \href{http://dx.doi.org/10.1103/PhysRevC.11.1}{{\em Phys. Rev. C} {\bfseries 11} (Jan, 1975) 1--18}. \url{https://link.aps.org/doi/10.1103/PhysRevC.11.1}.

\bibitem{Coester:NNN}
W.~Gl\"ockle, T.-S.~H. Lee, and F.~Coester, ``Relativistic effects in three-body bound states,'' \href{http://dx.doi.org/10.1103/PhysRevC.33.709}{{\em Phys. Rev. C} {\bfseries 33} (Feb, 1986) 709--716}. \url{https://link.aps.org/doi/10.1103/PhysRevC.33.709}.

\bibitem{PhysRevC.42.1965}
R.~Brockmann and R.~Machleidt, ``{Relativistic nuclear structure. I. Nuclear matter},'' \href{http://dx.doi.org/10.1103/PhysRevC.42.1965}{{\em Phys. Rev. C} {\bfseries 42} (1990) 1965--1980}.

\bibitem{Brockmann:1996xy}
R.~Brockmann and R.~Machleidt, ``{The Dirac-Brueckner Approach},'' \href{http://dx.doi.org/10.1142/9789812817501_0002}{{\em Int. Rev. Nucl. Phys.} {\bfseries 8} (1999) 121--169}, \href{http://arxiv.org/abs/nucl-th/9612004}{{\ttfamily arXiv:nucl-th/9612004}}.

\bibitem{Forest_Arriaga}
J.~L. Forest, V.~R. Pandharipande, and A.~Arriaga, ``Quantum monte carlo studies of relativistic effects in light nuclei,'' \href{http://dx.doi.org/10.1103/PhysRevC.60.014002}{{\em Phys. Rev. C} {\bfseries 60} (Jun, 1999) 014002}. \url{https://link.aps.org/doi/10.1103/PhysRevC.60.014002}.

\bibitem{boost}
J.~L. Forest, V.~R. Pandharipande, and J.~L. Friar {\em Phys. Rev. C} {\bfseries 52} (1995) 568.

\bibitem{Yang:2022esu}
Y.~L. Yang and P.~W. Zhao, ``{A consistent description of the relativistic effects and three-body interactions in atomic nuclei},'' \href{http://dx.doi.org/10.1016/j.physletb.2022.137587}{{\em Phys. Lett. B} {\bfseries 835} (2022) 137587}, \href{http://arxiv.org/abs/2206.13208}{{\ttfamily arXiv:2206.13208 [nucl-th]}}.

\bibitem{Yang:2024wsg}
Y.~L. Yang and P.~W. Zhao, ``{Reconciling light nuclei and nuclear matter: relativistic $ab\ initio$ calculations},'' \href{http://arxiv.org/abs/2405.04203}{{\ttfamily arXiv:2405.04203 [nucl-th]}}.

\bibitem{APR}
{A. Akmal, V.R. Pandharipande, and D.G. Ravenhall}, ``Equation of state of nucleon matter and neutron star structure,'' {\em Phys. Rev. C} {\bfseries 58} (1998) 1804.

\bibitem{AV18}
R.~B. Wiringa, V.~G.~J. Stoks, and R.~Schiavilla, ``{An Accurate nucleon-nucleon potential with charge independence breaking},'' {\em Phys. Rev. C} {\bfseries 51} (1995) 38--51.

\bibitem{UIX_2}
J.~Carlson, V.~R. Pandharipande, and R.~B. Wiringa {\em Nucl. Phys. A} {\bfseries 401} (1983) 59.

\bibitem{UIX_1}
B.~S. Pudliner, V.~R. Pandharipande, J.~Carlson, and R.~B. Wiringa, ``{Quantum Monte Carlo calculations of A <= 6 nuclei},'' {\em Phys. Rev. Lett.} {\bfseries 74} (1995) 4396--4399.

\bibitem{FF:CBF}
S.~Fantoni and A.~Fabrocini, ``{Correlated Basis Function Theory for Fermion Systems},'' in {\em {Microscopic Quantum Many-Body Theories and Their Applications}}, {J. Navarro and A. Polls}, ed., pp.~119--186.
\newblock {Springer, Berlin, Heidelberg}, 1998.
\newblock \url{https://doi.org/10.1007/BFb0104526}.

\bibitem{CLARK197989}
J.~W. Clark, ``Variational theory of nuclear matter,'' \href{http://dx.doi.org/https://doi.org/10.1016/0146-6410(79)90004-8}{{\em Progress in Particle and Nuclear Physics} {\bfseries 2} (1979) 89--199}. \url{https://www.sciencedirect.com/science/article/pii/0146641079900048}.

\bibitem{BL:2017}
{ O. Benhar and A. Lovato}, ``Perturbation theory of nuclear matter with a microscopic effective interaction,'' {\em Phys. Rev. C} {\bfseries 96} (2017) 054301.

\bibitem{Benhar_Lovato_Camelio}
O.~Benhar, A.~Lovato, and G.~Camelio, ``Modeling neutron star matter in the age of multimessenger astrophysics,'' \href{http://dx.doi.org/10.3847/1538-4357/ac8e61}{{\em The Astrophysical Journal} {\bfseries 939} no.~1, (Nov, 2022) 52}. \url{https://dx.doi.org/10.3847/1538-4357/ac8e61}.

\bibitem{lucas_temperature}
L.~Tonetto and O.~Benhar, ``Thermal effects on nuclear matter properties,'' \href{http://dx.doi.org/10.1103/PhysRevD.106.103020}{{\em Phys. Rev. D} {\bfseries 106} (Nov, 2022) 103020}. \url{https://link.aps.org/doi/10.1103/PhysRevD.106.103020}.

\bibitem{Benhar:Universe}
O.~Benhar, A.~Lovato, and L.~Tonetto, ``{Properties of Hot Nuclear Matter},'' \href{http://dx.doi.org/10.3390/universe9080345}{{\em Universe} {\bfseries 9} no.~8, (2023) 345}.

\bibitem{Friar:3BF}
J.~L. Friar, \href{http://dx.doi.org/10.1007/978-1-4684-5200-6_6}{``Trinucleon bound states,''} in {\em New Vistas in Electro-Nuclear Physics}, E.~L. Tomusiak, H.~S. Caplan, and E.~T. Dressler, eds., p.~213.
\newblock Springer US, Boston, MA, 1986.
\newblock \url{https://doi.org/10.1007/978-1-4684-5200-6_6}.

\bibitem{Fujita}
J.~Fujita and H.~Miyazawa, ``{Pion Theory of Three-Body Forces},'' {\em Prog. Theor. Phys.} {\bfseries 17} (1957) 360--365.

\bibitem{V6P}
R.~B. Wiringa and S.~C. Pieper, ``{Evolution of nuclear spectra with nuclear forces},'' \href{http://dx.doi.org/10.1103/PhysRevLett.89.182501}{{\em Phys. Rev. Lett.} {\bfseries 89} (2002) 182501}.

\bibitem{Nijmegen1}
J.~R. Bergervoet, P.~C. van Campen, R.~A.~M. Klomp, J.-L. de~Kok, T.~A. Rijken, V.~G.~J. Stoks, and J.~J. de~Swart, ``Phase shift analysis of all proton-proton scattering data below ${\mathit{t}}_{\mathrm{lab}}$=350 mev,'' \href{http://dx.doi.org/10.1103/PhysRevC.41.1435}{{\em Phys. Rev. C} {\bfseries 41} (Apr, 1990) 1435--1452}.

\bibitem{Nijmegen2}
V.~G.~J. Stoks, R.~A.~M. Klomp, M.~C.~M. Rentmeester, and J.~J. de~Swart, ``Partial-wave analysis of all nucleon-nucleon scattering data below 350 mev,'' \href{http://dx.doi.org/10.1103/PhysRevC.48.792}{{\em Phys. Rev. C} {\bfseries 48} (Aug, 1993) 792--815}.

\bibitem{SAID}
R.~A. Arndt, W.~J. Briscoe, I.~I. Strakovsky, and R.~L. Workman {\em Phys. Rev. C} {\bfseries 76} (2007) 025209.

\bibitem{Piarulli_benchmark}
M.~Piarulli, I.~Bombaci, D.~Logoteta, A.~Lovato, and R.~B. Wiringa, ``Benchmark calculations of pure neutron matter with realistic nucleon-nucleon interactions,'' \href{http://dx.doi.org/10.1103/PhysRevC.101.045801}{{\em Phys. Rev. C} {\bfseries 101} (Apr, 2020) 045801}. \url{https://link.aps.org/doi/10.1103/PhysRevC.101.045801}.

\bibitem{Foldy-Krajcik}
R.~A. Krajcik and L.~L. Foldy, ``Relativistic center-of-mass variables for composite systems with arbitrary internal interactions,'' \href{http://dx.doi.org/10.1103/PhysRevD.10.1777}{{\em Phys. Rev. D} {\bfseries 10} (Sep, 1974) 1777--1795}. \url{https://link.aps.org/doi/10.1103/PhysRevD.10.1777}.

\bibitem{Friar75}
J.~L. Friar, ``Relativistic effects on the wave function of a moving system,'' \href{http://dx.doi.org/https://doi.org/10.1103/PhysRevC.12.695}{{\em Phys. Rev. C} {\bfseries 12} (1975) 695--698}. \url{https://journals.aps.org/prc/abstract/10.1103/PhysRevC.12.695}.

\bibitem{Foldy1}
L.~L. Foldy, ``Relativistic particle systems with interaction,'' \href{http://dx.doi.org/10.1103/PhysRev.122.275}{{\em Phys. Rev.} {\bfseries 122} (Apr, 1961) 275--288}. \url{https://link.aps.org/doi/10.1103/PhysRev.122.275}.

\bibitem{CPS}
J.~Carlson, V.~R. Pandharipande, and R.~Schiavilla, ``Variational monte carlo calculations of $^{3}\mathrm{H}$ and $^{4}\mathrm{He}$ with a relativistic hamiltonian,'' \href{http://dx.doi.org/10.1103/PhysRevC.47.484}{{\em Phys. Rev. C} {\bfseries 47} (Feb, 1993) 484--497}. \url{https://link.aps.org/doi/10.1103/PhysRevC.47.484}.

\bibitem{Cowell:2002bv}
S.~T. Cowell and V.~R. Pandharipande, ``{Quenching of weak interactions in nucleon matter},'' \href{http://dx.doi.org/10.1103/PhysRevC.67.035504}{{\em Phys. Rev. C} {\bfseries 67} (2003) 035504}, \href{http://arxiv.org/abs/nucl-th/0211013}{{\ttfamily arXiv:nucl-th/0211013}}.

\bibitem{LAGARIS1981331}
I.~Lagaris and V.~Pandharipande, ``Phenomenological two-nucleon interaction operator,'' \href{http://dx.doi.org/https://doi.org/10.1016/0375-9474(81)90240-2}{{\em Nuclear Physics A} {\bfseries 359} no.~2, (1981) 331--348}. \url{https://www.sciencedirect.com/science/article/pii/0375947481902402}.

\bibitem{Akmal:1997}
A.~Akmal and V.~R. Pandharipande, ``Spin-isospin structure and pion condensation in nucleon matter,'' \href{http://dx.doi.org/10.1103/PhysRevC.56.2261}{{\em Phys. Rev. C} {\bfseries 56} (Oct, 1997) 2261--2279}. \url{https://link.aps.org/doi/10.1103/PhysRevC.56.2261}.

\bibitem{Maselli:2020uol}
A.~Maselli, A.~Sabatucci, and O.~Benhar, ``{Constraining three-nucleon forces with multimessenger data},'' \href{http://dx.doi.org/10.1103/PhysRevC.103.065804}{{\em Phys. Rev. C} {\bfseries 103} no.~6, (2021) 065804}, \href{http://arxiv.org/abs/2010.03581}{{\ttfamily arXiv:2010.03581 [astro-ph.HE]}}.

\bibitem{Sabatucci2022}
A.~Sabatucci, O.~Benhar, A.~Maselli, and C.~Pacilio, ``Sensitivity of neutron star observations to three-nucleon forces,'' \href{http://dx.doi.org/10.1103/PhysRevD.106.083010}{{\em Phys. Rev. D} {\bfseries 106} (Oct, 2022) 083010}. \url{https://link.aps.org/doi/10.1103/PhysRevD.106.083010}.

\bibitem{LucaG}
A.~Nasoni, E.~Filandri, and L.~Girlanda, ``{Relativistic constraints on 3N contact interactions},'' \href{http://dx.doi.org/10.1140/epja/s10050-023-01185-3}{{\em Eur. Phys. J. A} {\bfseries 59} no.~12, (2023) 293}, \href{http://arxiv.org/abs/2308.13341}{{\ttfamily arXiv:2308.13341 [nucl-th]}}.

\bibitem{Riley:2021pdl}
T.~E. Riley {\em et~al.}, ``{A NICER View of the Massive Pulsar PSR J0740+6620 Informed by Radio Timing and XMM-Newton Spectroscopy},'' \href{http://dx.doi.org/10.3847/2041-8213/ac0a81}{{\em Astrophys. J. Lett.} {\bfseries 918} no.~2, (2021) L27}, \href{http://arxiv.org/abs/2105.06980}{{\ttfamily arXiv:2105.06980 [astro-ph.HE]}}.

\bibitem{Riley:2019yda}
T.~E. Riley {\em et~al.}, ``{A $NICER$ View of PSR J0030+0451: Millisecond Pulsar Parameter Estimation},'' \href{http://dx.doi.org/10.3847/2041-8213/ab481c}{{\em Astrophys. J. Lett.} {\bfseries 887} no.~1, (2019) L21}, \href{http://arxiv.org/abs/1912.05702}{{\ttfamily arXiv:1912.05702 [astro-ph.HE]}}.

\bibitem{bob_vijay_rmp}
V.~R. Pandharipande and R.~B. Wiringa, ``{Variations on a theme of nuclear matter},'' \href{http://dx.doi.org/10.1103/RevModPhys.51.821}{{\em Rev. Mod. Phys.} {\bfseries 51} (1979) 821--859}.

\bibitem{forest_thomas}
J.~L. Forest, V.~R. Pandharipande, J.~Carlson, and R.~Schiavilla, ``Variational monte carlo calculations of $^{3}\mathrm{H}$ and $^{4}\mathrm{He}$ with a relativistic hamiltonian,'' \href{http://dx.doi.org/10.1103/PhysRevC.52.576}{{\em Phys. Rev. C} {\bfseries 52} (Aug, 1995) 576--577}. \url{https://link.aps.org/doi/10.1103/PhysRevC.52.576}.


\end{thebibliography}\endgroup

\end{document}